\RequirePackage{lineno}
\documentclass[preprint,preprintnumbers,amsmath,nofootinbib,amssymb,aps]{revtex4}
\usepackage{graphicx}% Include figure files
\usepackage{dcolumn}% Align table columns on decimal point
\usepackage{bm}% bold math
\usepackage{slashed}
\usepackage{ulem}
\usepackage{hyperref}
\usepackage{xcolor}
\colorlet{myPurple}{blue!40!red}
\newcommand\redsout{\bgroup\markoverwith{\textcolor{red}{\rule[0.5ex]{2pt}{0.4pt}}}\ULon}
\begin{document}
%hayrmer
%\setpagewiselinenumbers
%\modulolinenumbers[1]
%\linenumbers

\title{Towards  observation of three-nucleon short-range correlations in high $Q^2$  $A(e, e^\prime)X$ reactions}

\author{Donal B. Day$^{1}$,  Leonid L. Frankfurt$^{2}$, Misak  M. Sargsian$^{3}$  and Mark I. Strikman$^{4}$ }
\affiliation{
$^1$ Department of Physics, University of Virginia, Charlottesville, VA 22904, USA\\
$^2$Sackler School of Exact Sciences, Tel Aviv University, Tel Aviv, 69978, Israel \\
$^3$ Department of Physics, Florida International University, Miami, FL 33199, USA\\
$^4$ Department of Physics, Pennsylvania State University, University Park, PA 16802, USA}

\date{\today}

\begin{abstract}
We present a detailed study of  the kinematical and dynamical conditions necessary for probing highly elusive 
three-nucleon short range correlations~(3N-SRCs) in nuclei through inclusive electron scattering. 
The kinematic requirements that should be satisfied in order to isolate 3N-SRCs in 
inclusive processes are derived.  
We demonstrate that a  sequence of  two short-range NN interactions is 
the main 
mechanism for 3N-SRCs in 
such  processes.
With this mechanism we predict a quadratic dependence of the cross section ratios of nuclei to $^3$He 
in the 3N-SRC region to the same ratio measured in 2N-SRC domain. 
An  extended analysis of the available data which satisfies  the required 3N-SRC kinematical conditions is presented. 
This analysis reveals
 tantalizing  signatures of the scaling associated with the onset of   3N-SRC dominance.  
The same data  are also  consistent  with the prediction  of the quadratic  relation between the ratios 
measured in the 3N- and 2N-SRC regions for 
nuclei ranging $4 \le A \le 197$.  
This agreement made it possible to  extract  $a_3(A)$,  the probability of 3N-SRCs relative to  the A=3 nucleus. 
 We find $a_3(A)$ to be significantly larger in magnitude than the analogous parameter, $a_2(A)$, 
for 2N-SRCs.
\end{abstract}
\maketitle

%%%%%%%%%%%%%%%%%%%%%%%%%%%%%%%%%%%%%%%%%%%%%%%%%%%%%%%%%%%%%%%%%%%%%%%%%
\section{Introduction} 
\label{sec:intro}

For the last several  decades there have been intensive studies of two-nucleon short range correlations (SRCs) in nuclei 
using both electron\cite{FSDS,Kim1,Kim2,eip4,eip3,Fomin2011,srcprog} and proton\cite{eip1,eip2} probes in
high momentum transfer reactions.
Inclusive electron scattering experiments\cite{Kim1,Kim2,Fomin2011}, in which only the scattered electron is  detected, established the existence of the scaling properties associated with 2N-SRCs, confirming the observation\cite{FSDS}  based on the   analysis  of  
SLAC data\cite{Rock:1982gf,Arnold:1988us,Day:1979bx,Day:1987az,Schutz:1976he,Rock:1981aa}.
From these experiments $a_2(A,Z)$, describing  the probability of finding 2N-SRC in a given nucleus relative to the deuteron, was extracted.
The analysis of $A(p,2p)X$ data  yielded  similar estimates for  $a_2(A,Z)$ for proton induced reactions\cite{Yaron:2002nv}. 
Moreover, the strength of 2N-SRCs found in these analyses agreed with the one obtained from  fast backward nucleon production in high energy inclusive $p(\gamma) - A$ scattering \cite{FS81}. 
The consistency among these 
measurements of  the  2N-SRC strength with different probes
 supports the conjecture that a genuine property of the nuclear ground state wave function has been probed. 

The extension of 2N-SRC studies  to  semi-inclusive processes  in which, 
in addition to the scattered probe, the struck nucleons\cite{eip4,eip3} or both struck and recoil nucleons from 2N-SRCs\cite{eip1,eip2}
have been detected, discovered the strong (a factor of 20) dominance\cite{isosrc,eip3} of 
$pn$ SRCs as compared to $pp$ and $nn$ SRCs in the probed internal momentum range of $300-650$~MeV/c. The $pn$ excess is understood\cite{eheppn2,Schiavilla:2006xx} 
when considering the dominance of the tensor interaction at inter-nucleon distances of $0.8 - 1.2$~fm and which  supports 
the  commanding role of 2N-SRCs in the high momentum component of  the nuclear wave function. Based on the $pn$-SRC dominance 
it was predicted that minority component in asymmetric nuclei should have larger kinetics energy\cite{newprops} which was 
confirmed  experimentally\cite{twoferm,CLAS:2018xvc,CLAS:2018yvt}.

The  experimental  focus  on 2N-SRCs stimulated extensive  theoretical efforts (see e.g. 
Refs.\cite{FS81,CiofidegliAtti:1991mm,CiofidegliAtti:1995qe,Vanhalst:2014cqa,multisrc,CiofidegliAtti:2017tnm}) 
 to calculate  the   multitude of nuclear quantities  entering into the cross sections of inclusive and semi-inclusive 
 electron nuclear scattering. Such quantities are the nuclear spectral and decay functions  which are  based on the 
 2N-SRC model of    high momentum component of the  nuclear ground state wave  function.

A question which naturally arises is what is the structure of nuclear wave function at even  larger internal momenta of 
the bound nucleons ($> 650$~MeV/c). 
One of the important issues in this regard is the possible formation of 3N-SRCs. Understanding the strength and  dynamics of 3N-SRCs is essential to advance our knowledge of super-dense nuclear matter. In most realistic models of  the nuclear equation of state  3N-SRCs play an increasingly important role above the saturation densities (see e.g.  Ref.\cite{Heiselberg:2000dn}). 
The 3N-SRCs can be formed both by nuclear forces that can be reduced to a sequence of two short range elastic NN interactions
and by irreducible 3N forces that contain inelastic transitions in the intermediate state.

Experimental evidence for  3N-SRCs is very limited. One of the main obstacles in isolating and probing 3N-SRCs is that they have a much reduced probability compared to 2N-SRCs. 
%As follows from 
Analysis of  Lippmann-Schwinger type equations for nuclear bound states\cite{srcrev,arnps},
strongly suggests 
%it is likely 
that 2N-SRCs dominate the momentum distribution for momenta larger than those characteristic of 2N-SRCs. 
Thus the study of 3N-SRCs requires the  consideration  of 
variables other  than just the momentum of the bound nucleon. 
One such parameter is $\alpha$\cite{FS81}, the light-cone~(LC) momentum fraction of the nucleus carried by the bound nucleon. 
In  collider kinematics  $\alpha$ is equal to the ratio of the nucleon longitudinal momentum to  the nucleus  momentum, scaled by A; 
such that in the case of equal partition of the nuclear momentum, $\alpha=1$. The condition that $\alpha> 2$  requires at least three nucleons to be in close proximity in order for a single  nucleon to carry more than two nucleon's momentum fraction. An early analysis of few nucleon SRCs\cite{FS81} in the backward production of protons  
with momenta $0.3 < p < 1.5$~GeV/c indicated that the scattering off 3N-SRCs begins to dominate the 2N-SRC contribution  starting at  $\alpha \simeq 1.6$, which we consider as a kinematic   threshold for isolating 3N-SRCs.

In inclusive A(e,e')X reactions it is expected that the dominance of 3N-SRCs will be revealed by the onset of another  plateau in the  ratios of 
per-nucleon cross sections of heavy to light nuclei at $x > 2$. However, observation of such a plateau has been elusive. 
One of the first attempts to isolate 3N-SRC at Bjorken $x > 2$ observed a possible plateau\cite{Kim2},  though subsequent measurements of  the ratio 
$\frac{3\sigma_{^4He}}{4\sigma_{^3He}}$ did not make that claim \cite{Fomin2011}.       The most recent  measurement\cite{Ye:2017mvo} of the inclusive cross section ratios  of $^4$He to $^3$He at $x >2$ and $1.5 < Q^2 < 1.9$~GeV$^2$ are largely in agreement with Ref.~\cite{Fomin2011} in that  no  plateau was observed.   This situation corroborated the suggestion\cite{Higinbotham:2014xna} 
that poor momentum resolution for the scattered electrons in the experiment of  Ref.\cite{Kim2}
allowed   events to migrate from smaller to larger $x$ bins  and was responsible for the appearance of the plateau at  $x>2$.

In the recent work\cite{Sargsian:2019joj} we reported the partial analysis of inclusive $A(e,e^\prime)X$ data utilizing the above discussed kinematic variable $\alpha$ for  ($\gtrsim 1.6$) region.  We demonstrated that the data in this domain show a tantalizing signature for 
another layer of scaling for the ratio of per-nucleon inclusive cross sections ${3\sigma(^4He)\over 4\sigma(^3He)}$.  The analysis of other 
nuclei indicated also an  agreement with the theoretical prediction of a quadratic proportionality  of $a_3(A)$  to the 
ratio of ${a_2(A)\over a_2(A=3)}$ measured in the 2N~SRC domain.
 
In the current  paper we present a detailed theoretical analysis of inclusive scattering at $\alpha >1$ kinematics and provide 
a theoretical foundation for the observation of  a  new layer of nuclear scaling at $\alpha > \alpha^0_{3N}$ as well 
as the expectation of a quadratic proportionality between the probabilities of 2N- and 3N- SRCs. We also present a  more complete analysis of the experimental data of Ref.\cite{Fomin2011,fominphd} using varied approaches to treat the poor quality of the cross section at large $x$, expected to be dominated by 3N SRCs.

In Sec.~\ref{sec2} we elaborate on the kinematics of 3N-SRCs using the variable $\alpha$ that 
characterizes the light-cone momentum fraction of the nucleus carried by the bound nucleon. By analyzing the decay function of the $^3$He nucleus we identify the 
dominant  mechanism forming 3N-SRCs in inclusive $eA$ scattering  and,  within this picture,  calculate the 
LC momentum fraction, $\alpha_{3N}$,  corresponding to   scattering from a nucleon in 3N-SRC.
This variable allows us to identify the  optimal kinematics for probing 3N-SRCs in  inclusive scattering.  
Sec.~\ref{sec3} discusses the dynamical origin of 3N-SRCs. Based on the model in 
which 3N-SRCs are generated through the two successive $pn$ short range interactions it is  predicted that the light-cone nuclear density matrix  which enters in $A(e,e^\prime)X$ cross section  is proportional to the convolution of two $pn$-SRC density functions.  
In Sec.~\ref{sec4} final state interactions in 
inclusive processes are considered as a potential source  masquerading  or destroying   3N-SRCs. Here we employ the important property of 
high energy small angle scattering where the quantity $\alpha$  is  approximately conserved in  rescattering processes.   
An experimental observable of 3N-SRCs in $A(e,e^\prime)X$ reactions 
is presented in Sec.~\ref{sec5}, where we also derive  the quadratic relation between the ratios of inclusive cross sections measured in the 
3N- and 2N~SRC regions.  Sec.~\ref{sec6} presents the  analysis of the  existing inclusive data in light of the  theoretical considerations 
presented in the previous sections.  In Sec.~\ref{sec7} we summarize our results and give an outlook on the perspective of unambiguous verification 
of 3N~SRCs.

\medskip

\section{Definition of 2N and 3N~SRCs}
\label{sec2}
In a non-relativistic formulation we define  a nucleon to be in  a 2N~SRC pair if its momentum exceeds the characteristic nuclear Fermi momentum, ($k_{F} \sim 250$~MeV/c) and is 
 almost completely balanced by the momentum of the correlated nucleon in the pair. 
 In the light-cone representation the requirement is that LC momentum fractions of the correlated nucleons $\alpha_1$ and $\alpha_2$ satisfy 
 conditions: $\alpha_i\ge 1.3$ or $\alpha_i \le 0.7$ for $i=1,2$ and $\alpha_1 + \alpha_2 \approx 2$.
There are also 2N-SRCs with $\alpha_i\sim 1$ and $p_\perp>0.3$~GeV/c, however they are not important for inclusive scattering 
at $x>1$.

For the nucleon in  a 3N~SRC we assume  again,  that its momentum significantly exceeds $k_{F}$, but in this case this momentum is balanced by two correlated nucleons each 
with momenta exceeding $k_F$.  As in the case of 2N-SRCs the center of mass momentum of the 3N~SRC is small, $p_{cm} \le k_F$. 
The description of 3N-SRCs in the LF representation corresponds to the situation in which $\alpha_i\ge 1.3$ or $\alpha_i \le 0.7$ for $i=1,2,3$ and $\alpha_1 + \alpha_2 + \alpha_3 \approx 3$.
Similar to 2N-SRC case,  some 3N correlations may correspond to the kinematics in which $\alpha_i\sim 1$ with nucleons 
having very large  transverse momenta. We do not discuss here such correlations since they contribute very little to $A(e,e^\prime)X$ reactions at $x>1$.

The complete  nuclear wave function should incorporate components related to 2N- and 3N-SRCs. 
However, the first principle calculation of a wave function  containing  these components
is currently 
impossible due to a poor understanding of strong interaction dynamics at short internucleon distances. 
Relativistic effects that become increasingly important at large momenta of nucleons involved in short range correlations are also an impediment.

In this respect the progress can be achieved by experimental studies of 3N-SRCs which are currently becoming 
more accessible  with the 12 GeV energy upgrade of Jefferson Lab.
One way of addressing  the problem of  experimentally isolating   3N-SRCs,  is a proper  identification of the experimentally determined 
variables that can unambiguously discriminate 3N- from 2N- SRCs. 
As  was mentioned in the introduction, the  relevant variable 
is the light-cone momentum fraction of the nucleus carried by the interacting  bound nucleon, $\alpha$, first suggested in Ref.\cite{FS81,FS88}. 
The $\alpha$ variable,  in the reference frame in which nucleus has a large momentum in $-z$ direction, is defined  as 
\footnote{This variable has a equivalence with Bjorken $x_{Bj}$ that describes the light-cone momentum fraction of the nucleon carried by a parton.}: 
\begin{equation} \alpha = A {E_N - k_{N,z}\over E_A - k_{A,z}}, 
\label{alpha} 
\end{equation} 
where ($E_A$, $k_{A,z}$) and ($E_N$, $k_{N,z}$) are the energy and longitudinal momentum of the nucleus and bound nucleon respectively in the noncovariant LF nuclear wave function.

It was first suggested in Ref.\cite{FS81} that, due to the short-range nature of nuclear forces,  when
\begin{equation} 
j-1 < \alpha < j \label{srcon}, \end{equation} 
where $j>2$,  the scattering from j-nucleon SRC from the nucleus will be ensured.   
However, the fact that the probability of a j-nucleon~SRC in finite nuclei is $\approx ({r_{NN}\over r_{AV}})^{3(j-1)}$ with a correlation length $r_{NN}\ll r_{AV}$, where $r_{AV}$ is the average internucleon distance,  suggests  that the transition from $j$  to the $j+1$ SRC should occur at somewhat smaller 
$\alpha\lesssim j$~\cite{FS81}.  The latter inequality means that 3N-SRCs begin to dominate  at $\alpha  \simeq   2$.

\medskip

 \subsection{ 2N~SRCs} 
 In 1993, guided by Eq.(\ref{srcon}),   we studied the  possibility of exposing 2N- SRCs in high $Q^2$ inclusive $A(e,e^\prime)X$ reactions\cite{FSDS} by identifying  
 the relevant light-cone momentum fraction $\alpha_{2N}$ for inclusive processes as:
 \begin{equation} 
 \alpha_{2N} = 2 - {q_- + 2m_N\over  2 m_N}\left( 1 + {\sqrt{W^2_{2N} - 4m_N^2}\over W_{2N}}\right), 
\label{alpha2Ndef}
 \end{equation}
where $q_- = q_0 - |{\bf q}|$ and $W_{2N}^2 = (q+2m_N)^2 = -Q^2 +4q_0m_N + 4m_N^2$, with $m_N$ the nucleon mass, 
$q_0$ and $\bf q$ representing energy and momentum transfer and $Q^2 = {\bf q}^2 - q_0^2$. 
Eq.(\ref{alpha2Ndef}
%This equation 
explicitly takes into account the recoil energy and momentum carried by the spectator nucleon in the 2N-SRC and ensures that solutions for $\alpha_{2N}$ exist only for $ x\le 2$, where 
$x = {Q^2\over 2m_N q_0}$ is the Bjorken variable.  Additionally, in the limit of large $Q^2$, $\alpha_{2N}\approx x$, 
and  the variable $x$ can replace $\alpha_{2N}$ for identification of 
2N SRCs in the large $Q^2$ limit.

One of the important advantages of the LF treatment is that the inclusive scattering
cross section  can be factorized into the electron-bound nucleon scattering cross section, $\sigma_{eN}$  and 
light cone density matrix, $\rho_{A}(\alpha_{2N})$,  in the following form\cite{FS81,MS01}: 
\begin{equation} 
\sigma_{eA} \ \approx  \ \sum\limits_{N}\sigma_{eN}\rho_{A}(\alpha_{2N}).
\label{sigma_2N} 
\end{equation}     
Note that within a non-relativistic framework no such simple factorization exists and the inclusive cross section is expressed  through  an integral over $p_m^\perp$ and $E_m$ of the 
convolution of $\sigma_{eN}$ and the nuclear spectral function, $S_A(p_m,E_m$), 
where $p_m$ and $E_m$ are the missing momentum and energy in the reaction 
(see e.g. Ref.~\cite{ggraphs,MS01}).  
The theoretical justification for the factorization in 
Eq.~(\ref{sigma_2N}) in the  high $Q^2$ limit,  is based on the validity of the closure approximation 
over the ``plus"-component, $p_+$, of the 4-momentum of the bound nucleon in LF the 
formalism (see e.g. Ref.~\cite{MS01}). The $p_+$-component in the LF formalism is analogous to the  
 missing energy $E_m$ and in the calculation of  $\sigma_{eN}$  it  is estimated  at the  average point, corresponding to the 
 2N-SRC at rest, $p_+\approx 2m_N - {m_N\over 2-\alpha_{2N}}$ .  
Based on Eqs.~(\ref{sigma_2N}) and (\ref{srcon})   we predicted\cite{FSDS}  that due to the dominance of 2N-SRC dynamics,  the per-nucleon ratios of inclusive cross sections of nuclei and the deuteron:
 \begin{equation} 
 a_2(A,Z) = {2 \sigma_{eA}(\alpha_{2N}, Q^2)\over A \sigma_{ed}(\alpha_{2N},Q^2)},
 \label{a2} 
 \end{equation}
should scale  with $\alpha_{2N}$  for  $1.3 < \alpha_{2N} <2$ and  $Q^2>1.5$~GeV$^2$, with the parameter $a_{2}(A,Z)$  representing  the  probability of finding a 2N-SRC in  nucleus, A  relative to the deuteron. Here,  the lower limit of $\alpha_{2N}$ corresponds  to the scattering off a bound nucleon with average momentum of $p \gtrsim 0.3~\mbox{GeV/c}$.

The analysis of the available  data\cite{FSDS} at that time from large $Q^2$ inclusive experiments at SLAC  was in agreement with  the prediction of the scaling in Eq.(\ref{a2}). Subsequent dedicated experiments\cite{Kim1,Kim2,Fomin2011} at JLab confirmed this prediction and obtained similar  estimates for the scaling parameter $a_2(A,Z)$  for the wide range of atomic nuclei, $A$ (see e.g. Fig.\ref{2NSRC-Fomin} and the related discussion in Sec.\ref{sec6}.).

\begin{figure}[ht]
\vspace{-0.2cm}
\centering\includegraphics[scale=0.5]{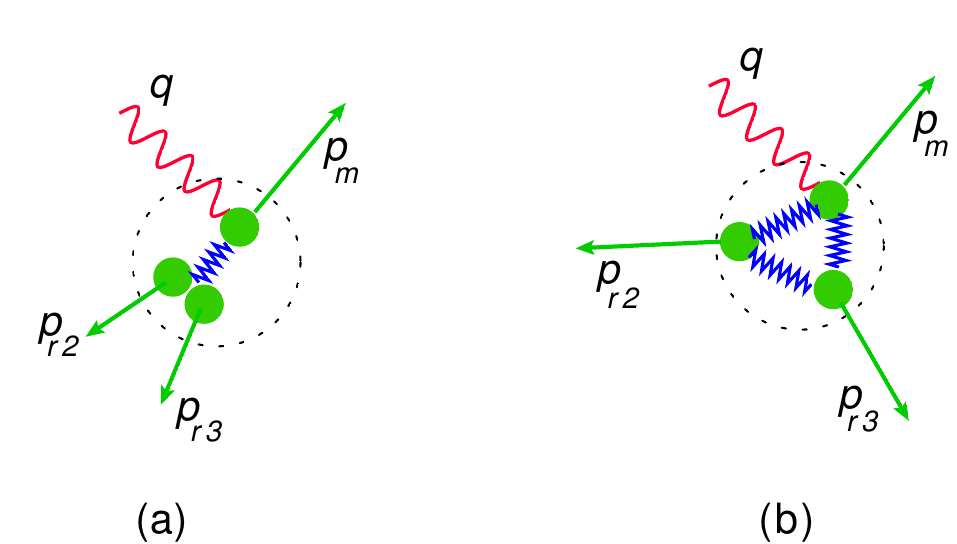}
\vspace{-0.2cm}
\caption{Types of three nucleon SRCs. (a) In type 3N-I SRC  the fast probed nucleon 
is balanced by two recoil nucleons with momenta $\approx p_m/2$. (b) In type 3N-II SRC 
all tree nucleons have equal momenta with relative angles $\approx 120^0$.}
\label{3Nsrc_types}
\end{figure}

\begin{figure}[ht]
\vspace{-0.2cm}
\centering\includegraphics[width=8.2cm,height=11cm]{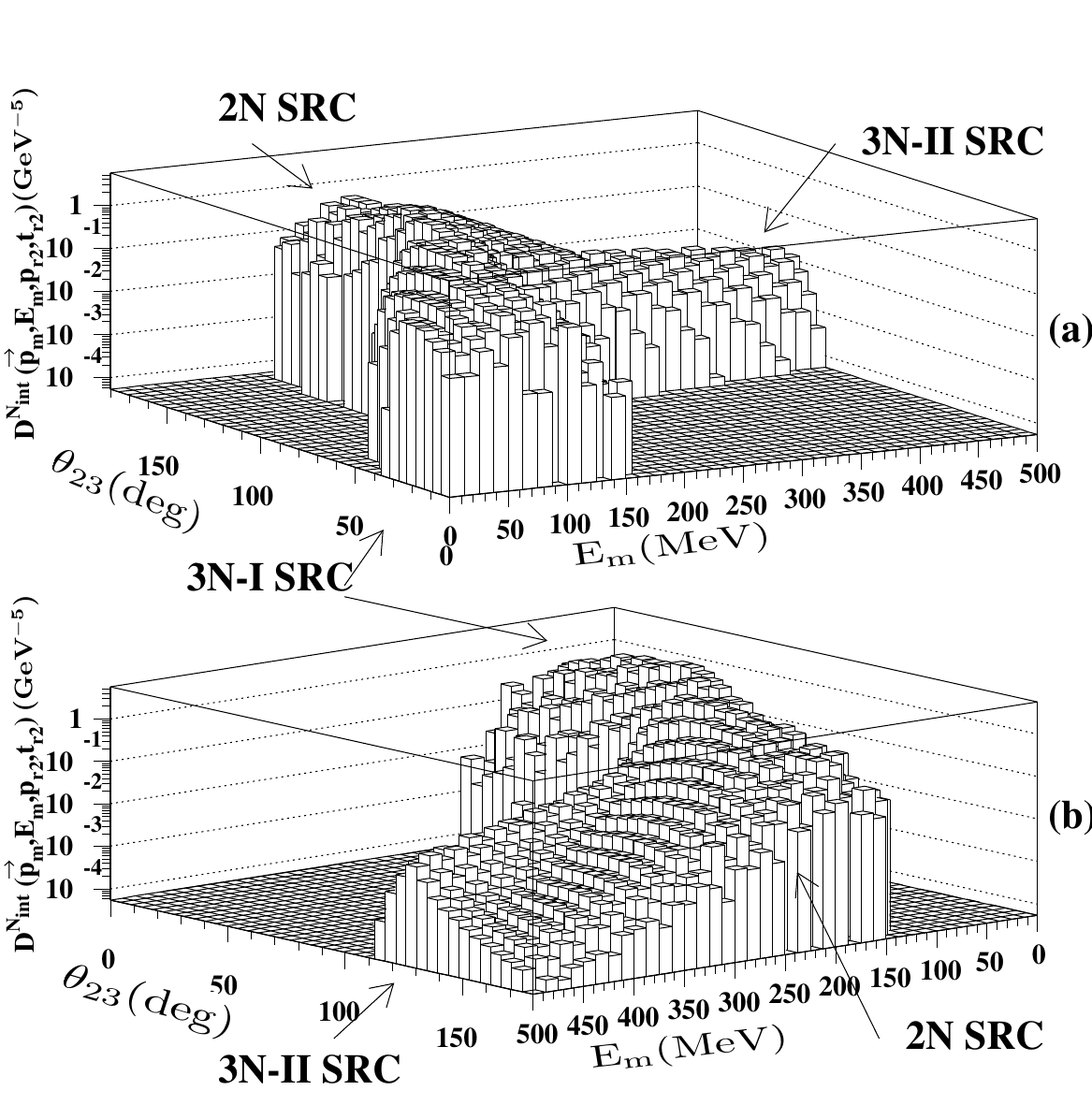}
\vspace{-0.4cm}
\caption{Decay function for $^3He$ nucleus calculated with the condition $p_m\ge 700$~MeV/c,  and $p_{r2},p_{r3}\ge k_{F}$.  The  $\theta_{23}$ is the relative angle between two recoil nucleons and $E_m$ is the missing energy. Two panels
show different point of views of the same figure. The figure is adapted from Ref.\cite{eheppn2}.}
\label{DecayFunction}
\end{figure}

\medskip

\subsection{3N~SRCs}

For 2N-SRCs we considered the only possible configuration in which two fast nucleons are correlated back-to-back with 
a small center of mass momentum. For 3N-SRCs, however there are more configurations in which three fast nucleons 
have 
%can be associated with a 
small center of mass momentum. Two extreme cases of possible 3N-SRC configurations are presented in 
Fig.~\ref{3Nsrc_types}.  The first, Fig.\ref{3Nsrc_types}(a),  referred as type 3N-I~SRC, corresponds to the situation in which the probed fast nucleon is balanced by two fast spectator nucleons $p_{r2},p_{r3}\sim p_m/2$  with a small relative angle between them, thus a small invariant mass, $m_S \sim 2 m_N$. The second  case, Fig.\ref{3Nsrc_types}(b) corresponds to the symmetric situation in which all three nucleons have comparable momenta with relative angles $\theta_{23} \sim 120^0$.   

To determine which of these  3N~SRC configurations will dominate in inclusive  $A(e,e^\prime)X$ scattering it is instructive to consider the decay function for a three-body nucleus at large values of missing and recoil momenta, noticing that it is  the integrated decay function  which enters in the cross section of  inclusive scattering. 
The decay function has been calculated in Ref.\cite{eheppn2,eheppn1} for $^3$He   using a realistic wave function based on the solution of Faddeev equations\cite{Nogga} and one  of the results relevant for 3N-SRCs is presented in Fig.~\ref{DecayFunction}. 
In the figure a correlation between  the relative angle of two recoil nucleons, $\theta_{23}$ and missing energy $E_m$ is presented
for $p_{m}\ge 700$~MeV/c and $p_{r2},p_{r3}> k_{F}$.   As the figure shows the type 3N-I SRC provides the  dominant contribution to the decay function at  small missing energies, $E_{m}\sim {p^2\over 4m_N}$, with the relative angle between spectator nucleons 
$\theta_{23}\le 50^0$ (see Fig.\ref{DecayFunction}(a)). 
A transition to the  type 3N-II SRC is observed with an increase  of  missing energy $E_{m}\ge  200$~MeV, in which case $\theta_{23}\sim 120^0$.  The analysis of type 3N-II SRCs\cite{eheppn2}  demonstrate that the irreducible three-nucleon forces have  substantial 
contribution in this region due to large missing energies which   increases the possibility of an inelastic $N\rightarrow \Delta$ transition at the NN vertices of the correlation.

Since the integrated  decay function, which enters in the  inclusive cross section,   is dominated by smaller values of $E_m$,
one expects, based on the above discussion, that the type 3N-I SRC  represents the main  configuration  
contributing  to the   inclusive cross section. 
Based on this, it is possible
to identify the kinematics at which 3N-SRCs can be isolated in inclusive scattering.
Introducing mass $m_S$ and momentum $p_S$ for the two nucleon recoil system of  type 3N-I SRC (Fig.\ref{3Nsrc_types}(a))   
we consider  energy-momentum conservation  in quasielastic scattering from a 3N-SRC which takes  the form:
\begin{equation}
q + 3m_N = p_f + p_{S}.
\end{equation}
Here $q$ is the four momentum transfer and $p_f$ is the final 4-momentum of the struck nucleon in the 3N~SRC.  
The boost invariance of the light-cone momentum fractions, for the spectator system in the   $\gamma- 3N$ center of mass frame, allows us to define the ratio:
\begin{equation}
{p^-_{s}\over p^-_{\gamma 3N}} = {E^{cm}_S + p^{cm}_{S,z}\over E^{cm}_S + E^{cm}_f} \approx {E^{cm}_S  + p^{cm}_{S}\over W_{3N}},
\label{psmn_g3nmn}
\end{equation}
where $W_{3N}$ is the invariant mass produced from the interaction with the 3N system:
\begin{equation}
W^2_{3N} = (q + 3m_N)^2 
= Q^2{3-x\over x} + 9 m_N^2.
\label{W2_3N}
\end{equation}
In the RHS  of  Eq.(\ref{psmn_g3nmn}) we neglected  the transverse momentum of the spectator 
NN system as it is integrated over in  inclusive reactions.  
This is justified since the inclusive cross section is dominated by kinematics in which $p_{S,\perp} \ll p_{S,z}$.
Furthermore  $E^{cm}_S$ and $p^{cm}_S$ can be calculated through $W_{3N}$ using the relation:

\begin{figure}[ht]
\vspace{-0.4cm}
\centering\includegraphics[width=9.2cm,height=14cm]{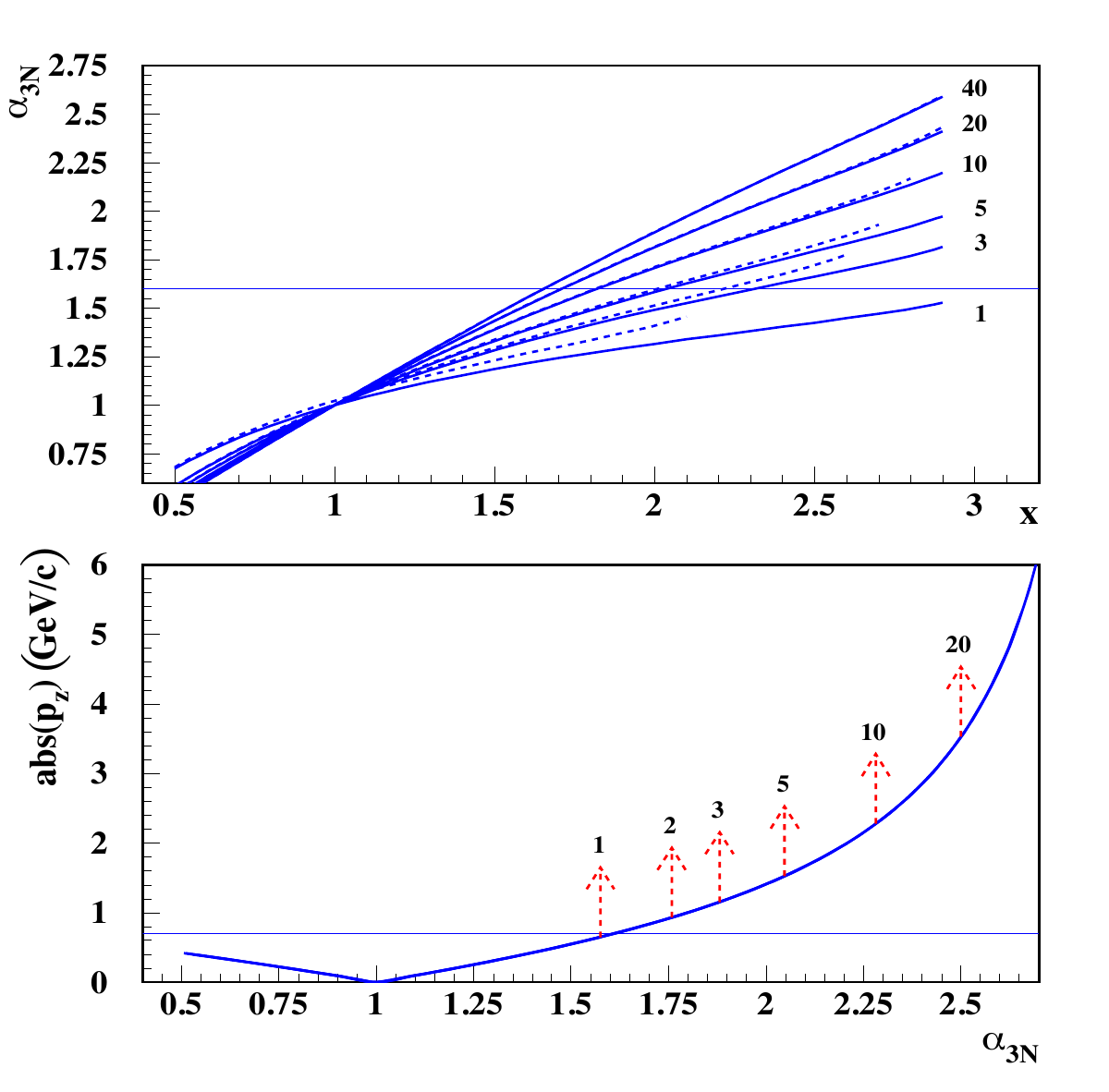}
\vspace{-0.8cm}
\caption{Kinematics of 3N~SRCs. (upper panel) Relation between $\alpha_{3N}$ and $x$ for $m_S$ calculated according to Eq.(\ref{mS})  with   $k=0$ (dotted line) and   $k = 250$~MeV/c.(dashed line). The curves are labeled by their respective 
$Q^2$ values.
(lower panel) The dependence of   $|p_z|$ on $\alpha_{3N}$. Arrows indicate the maximum possible $\alpha_{3N}$'s that can be reached at given values of $Q^2$.}
\label{3NSRCkin}
\end{figure}

\begin{equation}
E^{cm}_S  = {W_{3N}^2 - m_N^2 + m_S^2 \over 2 W_{3N}} \ \ \mbox{and} \ \ p^{cm}_S = \sqrt{E^{cm,2}_S - m_S^2},
\end{equation}
where  $m_S$ is defined as:
\begin{equation}
m_S^2  =   4 {m_N^2 + k_\perp^2 \over \beta(2-\beta)},
\label{mS}
\end{equation}
with  $k_\perp$ being the  transverse component of the relative momentum of the spectator nucleons with respect to $\vec p_S$. $\beta$ is the  light-cone momentum fraction of $p_S$ carried by one of the spectator nucleons and is scaled to be $0 \le \beta \le 2$.

Eq.(\ref{psmn_g3nmn}) can be used to estimate  the light-cone momentum fraction of the nucleon in a 3N-SRC by observing that 
$\alpha_{3N} = 3 - \alpha_{S}$, where $\alpha_S = 3{p^-_S\over   p^-_{3N}}$:
\begin{equation}
\alpha_{3N} = 3 -    3{p^-_S\over p^-_{3N}} =  3 -    3{p^-_S\over p^-_{\gamma 3N}}{p^-_{\gamma 3N} \over p^-_{3N}},
\end{equation}
where we again exploit the boost invariance of the ratio of ${p^-_{\gamma 3N} \over p^-_{3N}} = {q_- + 3m_N\over 3m_N}$ along ${\bf q}$.  This 
results in the following  expression for the light-cone momentum fraction of the nucleon with largest momentum belonging to
3N-SRCs:
\begin{eqnarray}
\alpha_{3N}   =  & & 3 \ - \ {q_- + 3 m_N\over 2 m_N} \left[1 \ \  + \ \ {m_S^2 - m_N^2\over W_{3N}^2} \ \ +  \right. \ \ \  \nonumber \\  
 & & \left.     
\sqrt{\left(1 - {(m_S + m_n)^2\over W_{3N}^2}\right)\left(1 - {(m_S - m_n)^2\over W_{3N}^2}\right)}\right].
\label{alpha3n}
\end{eqnarray}
Using this equation we can  identify the kinematical conditions 
for $x$ and $Q^2$  for which the inclusive cross section is dominated  by scattering from a nucleon in a 3N~SRC.     
For this, we first need to determine the threshold value, $\alpha^0_{3N}$,  above which one expects the onset of 
3N-SRC dynamics.  We also need  an estimate of $m_S$ through Eq.(\ref{mS}).
We note that for  inclusive $A(e,e^\prime)X$  scattering the cross section is defined by the nuclear light-cone density matrix where one integrates over the range of the two-nucleon spectator system masses $m_S \ge 2m_N$.  This integral however is dominated by $\beta \sim 1$ with the  recoil nucleon's  momentum, $k$  relative to $p_S$   
not exceeding  the nuclear Fermi momentum, $k_F \approx  250$~MeV/c (see e.g. Ref.\cite{eheppn2} ).   For numerical estimates 
we  consider two values for $k$:  $k = 0$ and $k = 250$~MeV/c.

With these considerations and  Eq.(\ref{alpha3n}) we are able to identify the  most favorable domain in $x$ and $Q^2$ 
to search for 3N-SRCs in inclusive $A(e,e^\prime)X$ reactions.  In Fig.\ref{3NSRCkin}(upper panel) we present the $\alpha_{3N}$ - $x$ relation for  different values of $Q^2$.  
The solid and dashed curves correspond to the spectator mass, $m_S$ calculated according to Eq.(\ref{mS})  with   $k=0$ 
and   $k = 250$~MeV/c.  
Figure~\ref{3NSRCkin} shows that at $Q^2\approx 3$~GeV$^2$ there  exists a finite kinematic  domain with $\alpha_{3N}\ge 1.6$ 
where one expects the onset of the 3N-SRC dominance.  In addition, starting with $Q^2\ge 5$~GeV$^2$ the onset of 3N-SRCs  is practically insensitive to the recoil mass of the spectator system, $m_S$.  As it follows from the figure, for  $Q^2 \gtrsim 3$ 
and $5$~GeV$^2$ the  magnitudes of $x\gtrsim 2.2$ and $x\gtrsim 2$  respectively are necessary to probe $\alpha_{3N}> 1.6$. 
Furthermore, using the relation, 
\begin{equation}
\alpha_S =   3 - \alpha_{3N} \approx   {\sqrt{m_S^2+p_z^2} + p_{z}\over m_N}, 
\label{alpha_S}
\end{equation}
we can calculate the longitudinal component of the initial momentum of the struck nucleon, $p_{z}$,  
which is the minimal possible momentum of the nucleon in 3N-SRC. 
According to  the 3N-SRC scenario  this momentum, $p_{z}$,  is equal and opposite to the center of mass momentum of the recoil two-nucleon system. It is worth mentioning that this momentum does not appear  directly in the argument of the light-cone nuclear wave function but  enters through the nonlinear relation of Eq.(\ref{alpha_S}). 
Nonetheless it gives an estimate of the  bound nucleon momenta  to be reached in a  fixed target experiment aimed at probing 3N~SRCs. 
Fig.\ref{3NSRCkin}(b) shows the dependence of   $|p_z|$ on $\alpha_{3N}$ with the arrows indicating the maximum possible 
$\alpha_{3N}$'s that can be probed at given values of $Q^2$. One observes  from the plot that the characteristic momenta of the struck nucleon  in the 3N~SRCs for $\alpha_{3N}\ge 1.6$ is $p_{z}\gtrsim 700$~MeV/c.

\section{Dynamics of the 3N~SRCs} 
\label{sec3}

\begin{figure}[ht]
\centering\includegraphics[scale=0.4]{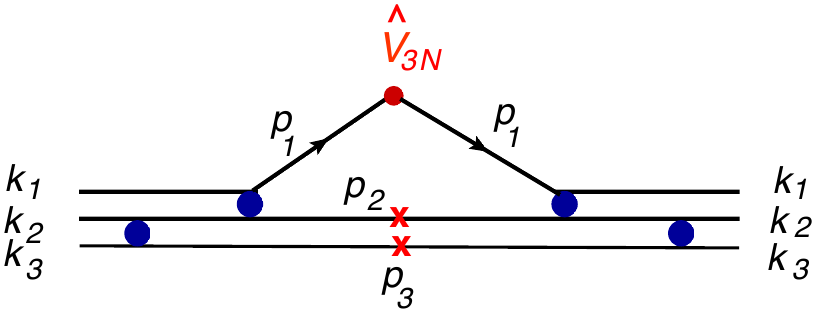}
\caption{The light-front diagram corresponding to the density matrix of 3N-SRCs. The 3N~SRCs  here are due to successive  $pn$ short range interactions. 
$k_i$ and $p_i$ are  shorthand notations for the light-cone momenta $(\beta_i,k_{i,\perp})$ and $(\alpha_i,p_{i,\perp})$. 
The figure is adapted from Ref.\cite{multisrc}.}
\label{3Nmodel}
\end{figure}

In light of the recent observation of strong dominance of  the $pn$ component  in
2N-SRCs\cite{isosrc,eip3,twoferm} within the momentum range of $250-650$MeV/c 
and the  assertion (discussed above) that type 3N-I SRCs dominate in inclusive 
scattering at $\alpha > 1.6$ and $\mid p_z\mid \gtrsim 0.7$~GeV/c one  expects that the 
main mechanism for  generation of 
3N-SRCs is due to successive  $pn$ short range interactions\cite{FS81,FSS15,multisrc} 
with the  mass of the spectator 2N system tending to 
be small, $m_S \sim 2m_N$.  Due to $pn$ dominance 3N-SRCs should have predominantly $ppn$ or $nnp$ composition
with $ppp$ and $nnn$  configurations being strongly disfavored. 
The diagram representing the   light-cone density matrix of 3N~SRCs is given in Fig.\ref{3Nmodel} where three-nucleon lines are  $ppn$ or $nnp$  configurations.
 
Calculation of the 3N-SRC contribution to the nuclear density matrix according to diagrams similar to  Fig.~\ref{3Nmodel} yields\cite{multisrc}:
\begin{eqnarray}
\rho_{3N}(\alpha_1)  & = &  \int {1\over 4}\left[ {3-\alpha_3\over (2-\alpha_3)^3}\rho_{pn}\left(\alpha_3,p_{3\perp}\right)
\rho_{pn}\left({2\alpha_2\over 3-\alpha_3}, p_{2\perp} + {\alpha_1\over 3-\alpha_3}p_{3\perp}\right)\right. + \nonumber \\
& & \left.
{3-\alpha_2\over (2-\alpha_2)^3}\rho_{pn}\left(\alpha_2,p_{2\perp}\right)
\rho_{pn}\left({2\alpha_3\over 3-\alpha_2}, p_{3\perp} + {\alpha_1\over 3-\alpha_2}p_{2\perp}\right)\right]
\delta(\sum\limits_{i=1}^3\alpha_i-3) \nonumber \\
& & d\alpha_2d^2p_{2\perp}
d\alpha_3d^2p_{3\perp},
\label{rho_3N}
\end{eqnarray}
where $(\alpha_i, p_{i\perp})$, ($i=1,2,3$) are light-cone momentum fractions and transverse momenta of 
nucleons and $\rho_{pn}(\alpha,p_\perp)$ is the density matrix of  $pn$-SRC.   
The prevalence of $\rho_{3N}$ in a nuclear density function, $\rho_A$,   in 3N-SRC region suggests several characteristics that can be 
experimentally verified. One follows from Eq.(\ref{a2}), according to which $\rho_{pn}\sim a_2(A,z)\rho_d$ and therefore 
the per nucleon probability of finding a nucleon in a  3N-SRC, $a_{3N}$, should be 
proportional to the square of the probabilities of  
2N~SRCs, $a_{2N}$,  (actual relation will be given in Sec.\ref{sec5}):
\begin{equation}
a_{3N}(A,Z) \sim a_{2N}(A,Z)^2.
\label{a3sima2a2}
\end{equation}

\begin{figure}[ht]
%\vspace{-2.4cm}
\centering\includegraphics[height=8cm,width=10cm]{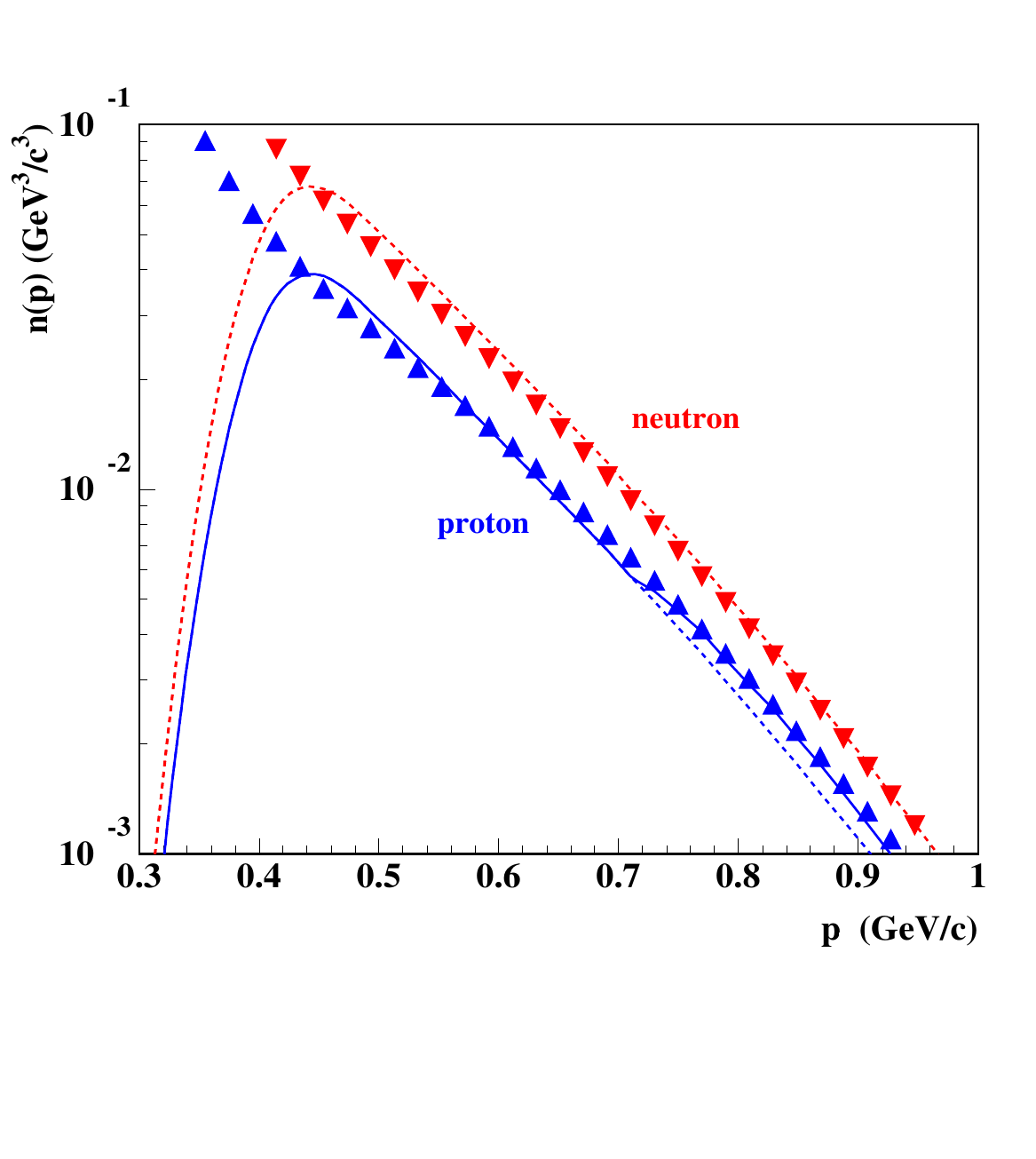}
\vspace{-1.2cm}
\caption{The momentum distribution of the proton and neutron in $^3$He. 
The triangle symbols are  from the VMC calculation of Ref.\cite{Wiringa}.  The dashed lines 
are contributions from 2N~SRCs only, solid lines  correspond to the combined contributions from 2N and 3N~SRCs\cite{multisrc}. 
In the case of the  neutron distribution  no 3N~SRCs are  included.}
\label{3hedist}
\end{figure}

Another  feature  follows  from the expectation that  the  mass of the recoil 2N system, $m_S$,  in 3N-SRC is small,
which results in a small relative momentum in the recoiling NN system, 
$k = {\sqrt{m_S^2 - 4m_N^2}\over 2}$. The condition $k\ll m_N$  and the fact that iso-triplet two-nucleon states with low relative momentum are strongly suppressed
compared to the iso-singlet states\cite{eheppn2}  produces a strong sensitivity of the 3N-SRCs on the isospin structure of NN recoil system.  
Namely, the dominant 3N-SRC configurations are those  which have a recoil 2N system in the iso-singlet state.  This situation is illustrated in 
Fig.~\ref{3hedist} where the high momentum distribution of protons and neutrons in $^3$He, calculated in a Variational Monte Carlo~(VMC) approach\cite{Wiringa}, 
is compared with the calculation based on  the 2N and 3N SRC model of Ref.~\cite{multisrc}, 
the latter being based on Eq.~(\ref{rho_3N}).
Fig.~\ref{3hedist} shows the 2N-SRC model completely describes the neutron momentum distribution  up to $1$~GeV/c, while
 one needs 3N-SRC contributions to describe the proton momentum distribution above $700$~MeV/c.  This result is in agreement with the 
 dominance of iso-singlet recoil NN systems in the  generation of 3N-SRCs.  
For the case of the neutron, the recoil system is a $pp$  pair,  which is  strongly suppressed as compared with that of
the proton, in which case the recoil system is in the  isosinglet $pn$ state where no suppression exists. 
Notice, that even if the  3N-SRCs contribute to the proton momentum distribution in $^3$He it is still a correction to the main 
2N-SRC part of the momentum distribution as  discussed in Sec.~\ref{sec:intro}.

It is worth mentioning that type 3N-II SRCs can be described through diagrams similar to Fig.~\ref{3Nmodel} in which 
case the intermediate state between two successive NN interactions has a  large invariant mass.  Here
another source of 3N-SRCs could be  the configuration containing a $\Delta$-resonance in the intermediate state,
which will represent the contribution from ``genuine" three-nucleon forces irreducible to NN interactions.  As it was discussed in Sec.\ref{sec2} one
expects that type 3N-I SRCs should be the dominant source of 3N correlations in inclusive reactions.  Probing type 3N-II SRCs will 
require semi-inclusive processes in which the  recoiling two-nucleon system has a large invariant mass.

\section{Final State Interactions}
\label{sec4}

Final state interactions~(FSI)  can both distort and mimic 3N~SRCs. 
Detailed quantitative studies of the FSI effects are clearly necessary. Below we provide several qualitative considerations based 
on the high energy nature of electro-production reactions which are used to probe 3N-SRCs. 
The main part here is that our interest is in LC momentum distribution function $\rho_(\alpha_{N})$, with $\alpha_N$  being the momentum 
fraction of nucleus  carried by the interacting nucleon. The latter is analogous to the Bjorken-x, which is in  deep-inelastic processes represents 
the LC momentum fraction of nucleon carried by the interacting quark. Our arguments related to the FSI are similar to that of DIS processes in which 
final state interaction of struck quark does not change the initial partonic distribution in the nucleon.  In this respect, it can be 
shown that ${\rho(\alpha)\over (2-\alpha)}$ is analogous to the partonic distribution function defined for the nucleon in the nucleus.

\begin{figure}[ht]
\centering\includegraphics[height=2.0cm,width=12.5cm]{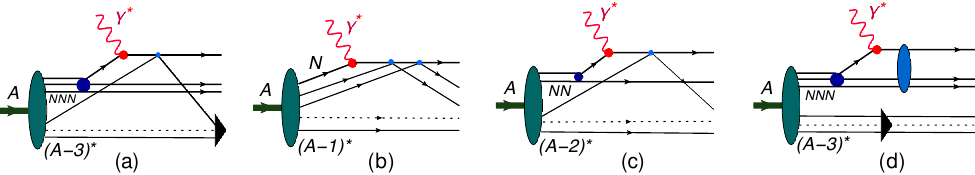}
\caption{Possible FSI diagrams contributing in 3N~SRC kinematics. Detailed description given in the text.}
\label{3N_FSI}
\end{figure}

The 3N-SRC picture  will be distorted  mainly due to the multiple rescattering 
of nucleons from 3N~SRCs  with  the nucleons belonging to the ``uncorrelated" spectator (A-3) system. 
An example is presented in Fig.\ref{3N_FSI}(a)  in which a  nucleon knocked-out from a 3N-SRC rescatters off the uncorrelated nucleons in the (A-3) 
residual nucleus.  Other examples are  the rescattering of spectator  nucleons in the  
3N-SRC with the  uncorrelated nucleons from (A-3) system.
For such rescatterings,  because of  inclusive nature of the process,  one integrates over the  range of excitation energies of the (A-3) system. 
As a result  the closure approximation can be applied (see discussion in Sec.\ref{sec2})  which cancels the effects of long-range 
FSIs. 
The empirical evidence for such  cancellation follows from the experimental observation of the 2N-SRC scaling in the $1 < x < 2$ 
domain\cite{FSDS,Kim1,Kim2,Fomin2011} (also Sec.\ref{sec2}) in which case the  closure condition is satisfied for the  FSI with  residual  (A-2) nucleus.  

FSIs  that can in principle mimic 3N-SRCs are diagrammatically presented  in Fig.\ref{3N_FSI}~(b) and (c). 
In the case of (b) an uncorrelated  nucleon in the mean field with initial LC momentum fraction,  $\alpha_N \approx 1$,  is struck  and two successive  rescatterings may increase the momentum faction to $\alpha_N  \gtrsim 1.6$, making it appear as a nucleon from 3N-SRC. 
In the case of (c) a nucleon  is knocked out from a 2N-SRC where  the characteristic momentum fraction  is $1.3\le \alpha_N \le 1.5$ and  
FSI shifts it   to the  $\alpha_N  \gtrsim 1.6$ region. 
An  important feature  that suppresses the migration of a nucleon of modest $\alpha_N$ into the 3N-SRC region is the approximate conservation of the  
LC momentum fraction in  high energy (eikonal) regime 
of  small angle rescattering\cite{ggraphs,MS01}.  
In this case the non-conservation of  $\alpha_N$ is estimated as\cite{ggraphs,MS01}:
\begin{equation}
\delta \alpha \approx {x^2\over Q^2}{2m_N E_R\over (1 + {4m_N^2x^2\over Q^2})},
\label{dalpha}
\end{equation}
where $E_R = \sqrt{m_S^2 + p^2} - m_S$ and $p\sim 0.7 - 1$~GeV/c  is the characteristic  momentum of the nucleon in a 
3N-SRC.

\begin{figure}[ht]
\vspace{-0.6cm}
\centering\includegraphics[height=7cm,width=8cm]{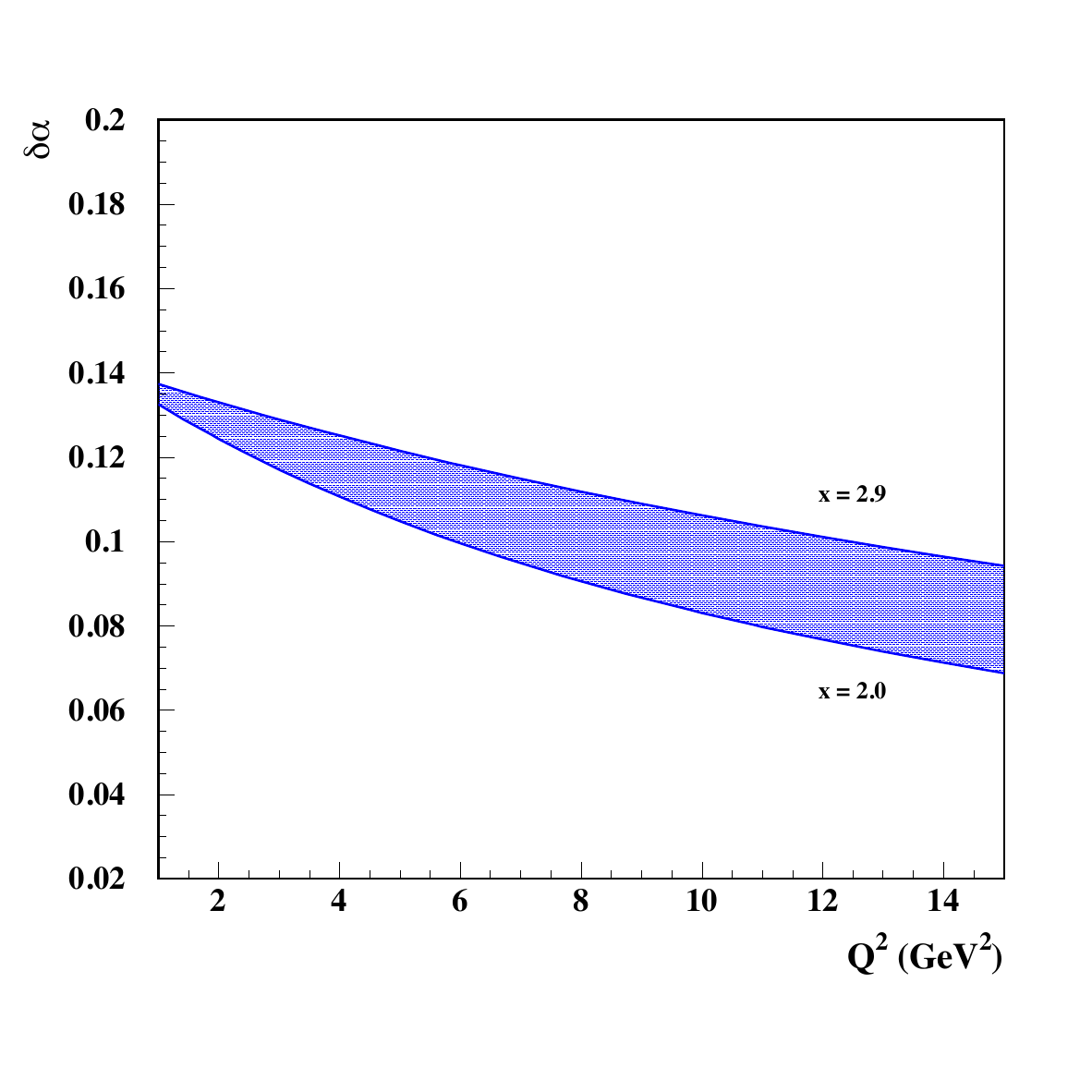}
\vspace{-0.6cm}
\caption{Non-conservation of  $\alpha_N$  as a function of $Q^2$ according to Eq.(\ref{dalpha}) for $p=1$~GeV/c 
and $2 \le x \le 2.9$.}
\label{delta_al}
\end{figure}

In Fig.~\ref{delta_al} we present the $Q^2$ dependence of the non-conservation of $\alpha_N$ for $2 \le x \le 2.9$. The estimates are made for $p=1$~GeV/c and $m_S = 2m_N$.  
It follows from the figure that  for $Q^2 > 3$~GeV$^2$ that FSI may alter  $\alpha_N$ by no more than $0.14$ which is too small to shift the mean field nucleon, $\alpha_N \approx 1$, to the 3N~SRC domain.  
However one may expect possible FSI contributions from the 2N-SRC domain, $1.3 \le \alpha_{2N} \le1.5$,   influencing the onset of 3N~SRCs at  $\alpha_{3N}\sim 1.6$.

Finally, the other FSI effects follow from the rescattering  within a 3N-SRC  as shown in Fig.\ref{3N_FSI}~(d). 
In this case one expects  a modification of the $p_\perp$ distribution in the 3N~SRCs. 
However the important  feature of high energy  small angle re-scattering, discussed above,  
is that while FSI redistributes  transverse momenta, it leaves  the  $\alpha_N$ distribution almost intact 
(see also Ref.\cite{dreview}). 
As a result the measured inclusive cross section in the 3N-SRC domain can be presented 
in the factorized form similar to the 2N-SRC case (Eq.(\ref{sigma_2N})):

\begin{equation}
\sigma_{eA}  \approx \sum\limits_{N} \sigma_{eN} \rho^N_A(\alpha_{3N}),
\label{crosssection}
\end{equation}
where 
\begin{equation}
\rho^N_A(\alpha_{3N}) = \int \rho^N_A(\alpha_{3N},p_\perp) d^2p_\perp,
\end{equation}
is weakly modified due to FSI, even if the $p_\perp$ distribution of the unintegrated nuclear density matrix,   
$\rho^N_A(\alpha,p_\perp)$ is distorted  by  the FSI~\cite{dreview}.

In the  discussion above we focused only on the part of the FSI which corresponds to the pole contribution from the struck nucleon propagator,  
representing 
the on-shell  propagation of the fast  nucleon in the intermediate state. 
Another contribution to  FSI comes from the non-pole term of the FSI amplitude in which case  the struck nucleon is highly  virtual.  
There are two main sources of the suppression of the off-shell FSI contribution. First,  the   off-shell FSI contribution is proportional to the square of the real part of the $NN$ scattering amplitude  
which is, smaller  by an order of magnitude,  than  the imaginary part (see e.g. \cite{Cosyn:2013uoa,Cosyn:2017ekf}). Second, due to the large virtuality of the fast nucleon 
the  off-shell $NN$ rescattering amplitude is strongly suppressed\cite{noredepn}. There is also an empirical evidence from the studies of 2N-SRCs that off-shell 
rescattering amplitudes are negligible\cite{FSDS}.

\section{3N~SRC Observables} 
\label{sec5}

The experimental observation of 3N-SRCs is challenging for many reasons. 
As Fig.~\ref{3hedist} shows,  that  extracting the  momentum distribution at $\gtrsim 700$~MeV/c 
will not allow the isolation of 3N-SRCs due to  substantial 2N-SRC contribution.  
Furthermore, the  3N-SRC  contribution to the momentum distribution  decreases faster  with 
an increase of momentum than the 2N-SRC contribution. 
Overall, the bound nucleon momentum is not a good parameter with which to explore 3N-SRCs.  
The more natural  parameter, as it was discussed earlier,
is the light-cone momentum fraction $\alpha_N$ for which according to Eq.(\ref{srcon}) the condition $\alpha_N\gtrsim 2$ will completely isolate 
3N-SRCs with the transition region expected to  start at $\alpha_N \gtrsim 1.6$.

 Moreover, according to Eq.(\ref{crosssection})  the cross section of inclusive reaction  factorizes in the form of the product of electron - nucleon cross section and  the 
$p_\perp$ integrated light-cone density matrix, $\rho_A(\alpha_{3N})$.
Hence, the appropriate  observable for  3N~SRCs is  the 
ratio of  inclusive $A(e,e^\prime)X$ cross sections for nuclei $A_2$ and $A_1$ 
in the region of $\alpha_{3N}\ge \alpha^0_{3N}$ and $Q^2> 3$~GeV$^2$:
\begin{equation}
R_{A_1}(A_2) = {A_1 \sigma_{A_2}(x,Q^2)\over A_2 \sigma_{A_1}(x,Q^2)}\left|_{\alpha_{3N}>\alpha^0_{3N}}\right. .
\label{R3nsrc}
\end{equation}
In this case  $\alpha^0_{3N}$ (expected to be $\approx 1.6$) 
should be defined from the observation of the onset of a {\em plateau}  
in the $\alpha_{3N}$ dependence of the ratio  $R_{A_1}(A_2)$.  Note that in Eq.(\ref{R3nsrc}) the  off-shell effects  
in electron-bound nucleon scattering are mostly cancelled in the ratio.

\begin{figure}[ht]
\centering\includegraphics[height=9cm,width=12cm]{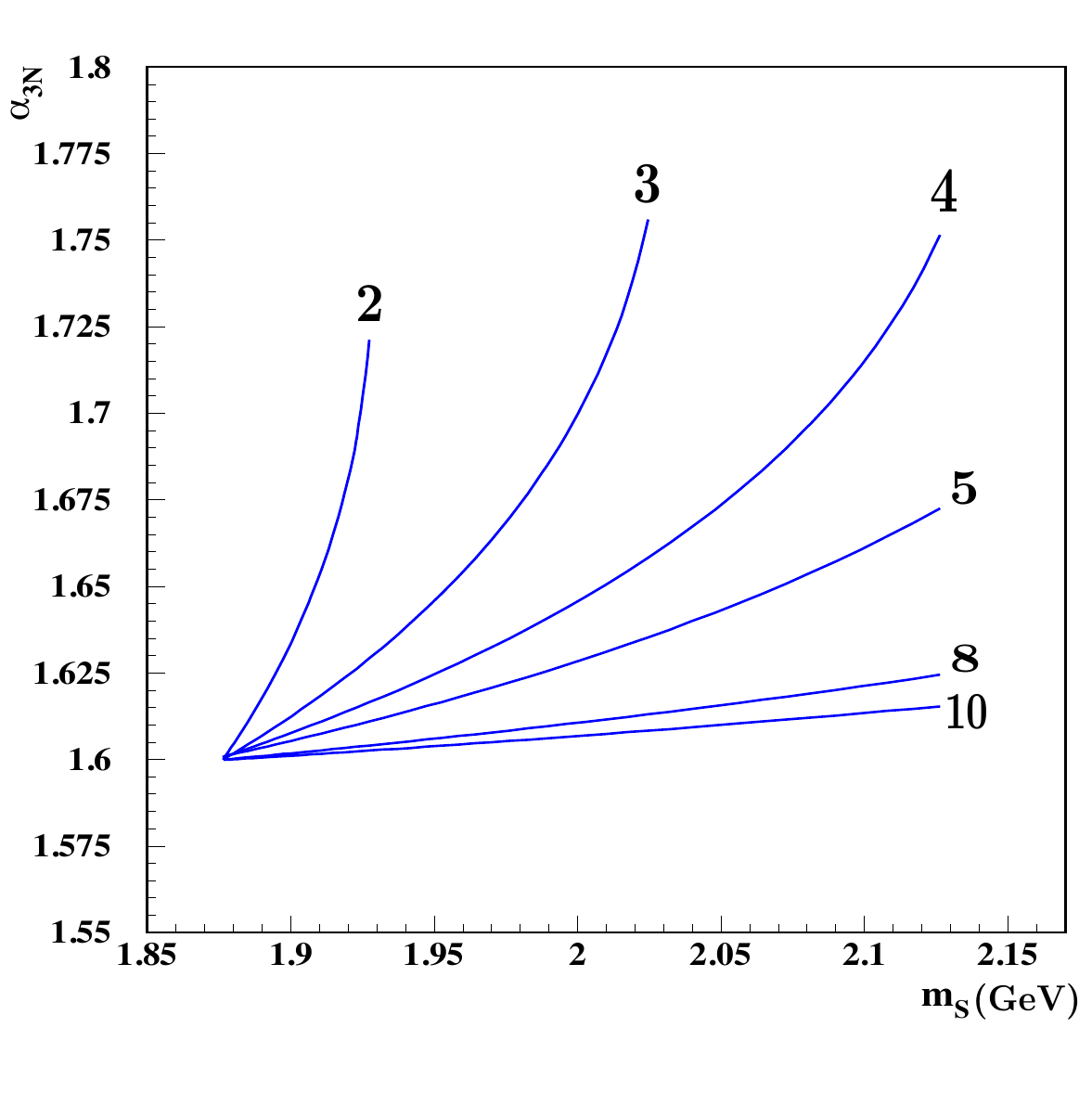}
\caption{Dependence of $\alpha_{3N}$ on the recoil mass, $m_s$  of the spectator system in 3N SRC for different values of $Q^2$,
calculated according to Eq.(\ref{alpha3n}).
}
\label{al3_ms}
\end{figure}

The observation of a plateau assumes also that $\alpha_{3N}$ is insensitive to the  recoil mass of the  
spectator 2N system, $m_S$,  over which the cross section of the inclusive scattering is integrated. 
This imposes an additional condition for the observation of scaling.  
This insensitivity is shown in Fig.~\ref{al3_ms}
and is largely  achieved at $Q^2\ge 5$~GeV$^2$.  
However  the expectation  that the integral over the recoil mass will saturate in the range of $2m_N \le m_S \le  2$~GeV\cite{eheppn2} 
indicates a possibility of  an early  onset of the plateau already at $Q^2=3$~GeV$^2$.

\begin{figure}[ht]
\centering\includegraphics[height=8cm,width=12cm]{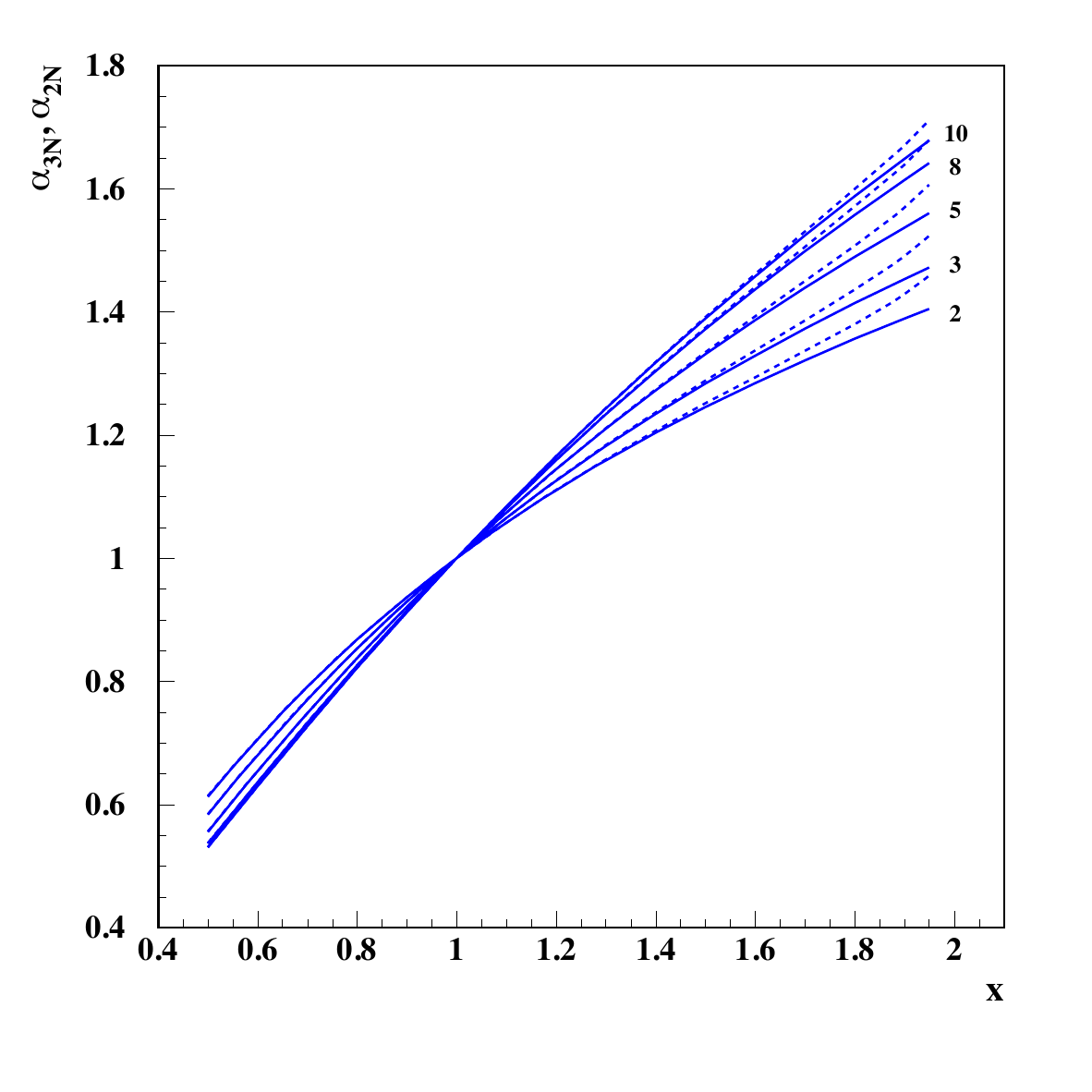}
\caption{The $x$ dependence of $\alpha_{3N}$ (solid lines) and $\alpha_{2N}$ (dashed lines) at different $Q^2$. }
\label{al_x}
\end{figure}

In the region of $\alpha_{3N} < \alpha^0_{3N}$  at modest $Q^2$ ($\le 3$~GeV$^2$) one  expects an existence of pre-asymptotic domain where  the ratios (\ref{R3nsrc})  are  not constant  as a function of  $\alpha_{3N}$ but  are largely  $Q^2$ independent  for fixed $\alpha_{3N}$.  This is mainly due to the factorization of  the inclusive cross section in the form of 
Eq.~(\ref{crosssection}).   
Such a  behavior  would be  analogous to  the pattern  observed for   $\alpha_{2N}$ dependence of  the ratio (\ref{a2})  
at  $\alpha_{2N} < 1.3$\cite{FSDS}. This is reinforced in Fig.~\ref{al_x} where one observes 
that $\alpha_{2N}$ and $\alpha_{3N}$ are nearly  identical for $x<1.6$.

To connect the ratio $R_{A_1}(A_2)$,  defined in Eq.(\ref{R3nsrc}),   with  theoretical calculations 
of nuclear density function we introduce parameter $a_3(A,Z)$ characterizing the probability of 3N SRCs 
for nearly symmetric nuclei as follows:
\begin{equation}
a_3(A,Z) = {3\over A} {\sigma_{eA}\over (\sigma_{e^3He} + \sigma_{e^3H})/2}.
\label{a3}
\end{equation}
This parameter can be related to the ratio $R_{3}(A,Z)$  which is defined in 
Eq.(\ref{R3nsrc})  for $A_2 = A$ and $A_1 = ^3$He. 
The ratio  $R_3(A,Z)$ is the most accessible experimental quantity.

Based on the factorization of Eq.(\ref{crosssection}), for $R_3(A,Z)$ and  $a_3(A,Z)$  one obtains:
\begin{equation}
R_3(A,Z) =  a_3(A,Z) {(\sigma_{ep}+\sigma_{en})/2\over (2\sigma_{ep} + \sigma_{en})/3}.
\label{R3_a3}
\end{equation}
Thus, a measurement of the ratio $R_3(A,Z)$ will allow an extraction of  the parameter $a_3(A,Z)$ which can be used for 
verification of the theoretical models of  3N SRCs.

Based on the above  definitions  we can also formulate  an  experimental observable  which  
will allow us to verify the prediction of Eq.(\ref{a3sima2a2}).  For this, we notice  that for type 3N-I SRC (Fig.\ref{3Nsrc_types}(a))  the calculation of nuclear density function\cite{multisrc} 
(Eq.(\ref{rho_3N}))  yields:
\begin{equation}
a_3(A,Z) = {a_2(A)^2\over a_2^p(^3He)a_2^n(^3He)},
\label{a3_l}
\end{equation}
where $a_2^p$ and $a_2^n$ are  per-nucleon probabilities of finding proton or neutron in a 2N- SRC.
One can relate these  to the parameter $a_2(A,Z)$ of Eq.(\ref{a2}) using the estimate of high momentum 
part of the proton and neutron distributions in nuclei  within  the $pn$-dominance model  in the form\cite{newprops}:
\begin{equation}
n^{p/n}_{2}(p) = {a_2(A)\over 2 (X^{p/n})^\gamma} n_d(p),
\label{pn_mod}
\end{equation}
where $X^{p/n} = {Z(N)\over A}$ is the relative fraction of the protons or neutrons and $n_d(p)$ is the high momentum distribution of the deuteron.
According to Eq.(\ref{pn_mod})  one estimates:
\begin{equation}
a_2^n(^3He) = {a_2(^3He)\over 2(1/3)^\gamma}  \ \ \ \mbox{and} \ \ \ a_2^p(^3He) = {a_2(^3He)\over 2(2/3)^\gamma},
\end{equation}
where $a_2(^3He)$ is defined according to Eq.(\ref{a2}). 

Using above estimates together with Eq.(\ref{R3_a3}) and Eq.(\ref{a3_l}) one obtains:   
\begin{eqnarray}
R_3(A,Z)   & = & 4\left({2\over 9}\right)^\gamma{(\sigma_{ep}+\sigma_{en})/2\over (2\sigma_{ep} + \sigma_{en})/3} \left( {a_2(A,Z)\over a_2(^3He)}\right)^2  = \nonumber \\
& &  4\left({2\over 9}\right)^\gamma{(\sigma_{ep}+\sigma_{en})/2\over (2\sigma_{ep} + \sigma_{en})/3} R_2^2(A,Z),
\label{R3_R2sq}
\end{eqnarray}
where in the last part of the equation we used the fact that 
the variables $\alpha_{2N}$ and $\alpha_{3N}$ have nearly same   magnitudes in the 2N-SRC region
(see in Fig.(\ref{al_x})) to 
relate the ratios of $a_2$ parameters to experimentally measured ratio:
\begin{equation}
R_2(A,Z) = {3\over A}{\sigma_{eA}\over \sigma_{e^3H}}\mid_{1.3 < \alpha_{3N}<1.5}  = {a_2(A)\over a_2(^3He)}.
\label{R2_def}
\end{equation}
In the following section we will analyze experimental data at $Q^2\sim$ 3~GeV$^2$ for which  $\sigma_{ep} \approx 3\sigma_{en}$.  
This and using $\gamma \approx 0.85$ from Ref.\cite{newprops} yields from  Eq.(\ref{R3_R2sq}):
\begin{equation}
R_3(A,Z) \approx 0.96R_2(A,Z)^2 \approx R_2(A,Z)^2.
\label{R3_R2sq_simp}
\end{equation}
Equations (\ref{R3_R2sq}) and (\ref{R3_R2sq_simp}) present a remarkable prediction: the ratios of inclusive nuclear cross sections ($R_2$ and $R_3$) measured 
in  different domains of $\alpha_{3N}$, will be related by simple quadratic relation if the scattering in the $\alpha_{3N}> \alpha^0_{3N}$ region probes 
type 3N-I SRCs.

\section{Experimental Evidence for 3N SRCs}
\label{sec6}

\begin{figure}[!htbp]
      \centerline{\includegraphics[width=.45\textwidth]{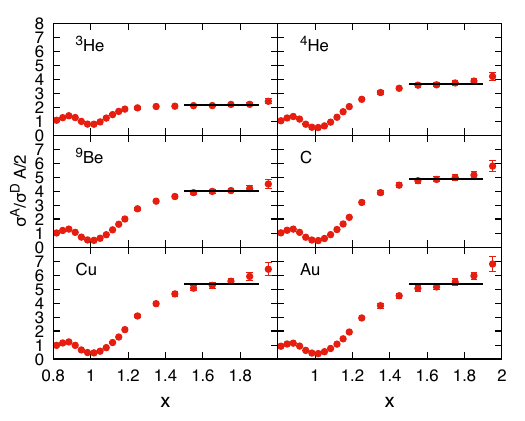}
      \includegraphics[width=.45\textwidth]{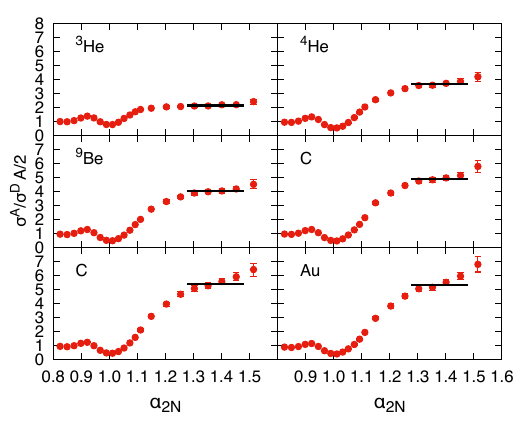}}
	\caption{Data from E02-019\cite{Fomin2011,fominphd} showing the ratios of ${2\sigma_A\over A \sigma_D}$ against $x$ and $\alpha_{2N}$ (right). The horizontal lines are the $a_2$ plateau values taken from \cite{Fomin2011}. 
	}
      \label{2NSRC-Fomin}
   \end{figure}

Conclusive evidence for two-nucleon SRCs first appeared in 1993\cite{FSDS}  from the  analysis of data from different experiments at  SLAC.  The SLAC data-sets for  light nuclei did not share common kinematics with the data for heavy nuclei\cite{Day:1993md} and it was necessary, after re-binning into common x-bins,  to interpolate the deuteron data across $Q^2$ to form the ratios of inclusive cross sections for nuclei $A$ and the deuteron  
(${2\sigma_A\over A\sigma_D}$). The plateau for the available nuclei in these ratios had a weak A dependence for A$\geq$ 12. The ratios were  smaller for $^3$He and $^4$He (with large error bars).  The  ${3\sigma_A\over A \sigma_{^3He}}$ ratios from  Hall B at JLab showed similar plateaus\cite{Kim1,Kim2}.    These  measurements  provided  persuasive evidence for the presence of 2N~SRCs yet were limited in  their precision and/or the desired expansive range in  $x$ and Q$^2$. Most recently, experiment E02-019\cite{Fomin2011,fominphd} produced high quality data in the 2N-SRC region  - these are reproduced in Fig.~\ref{2NSRC-Fomin}.

\begin{figure}[!htbp]
      \centerline{ \includegraphics[width=\textwidth]{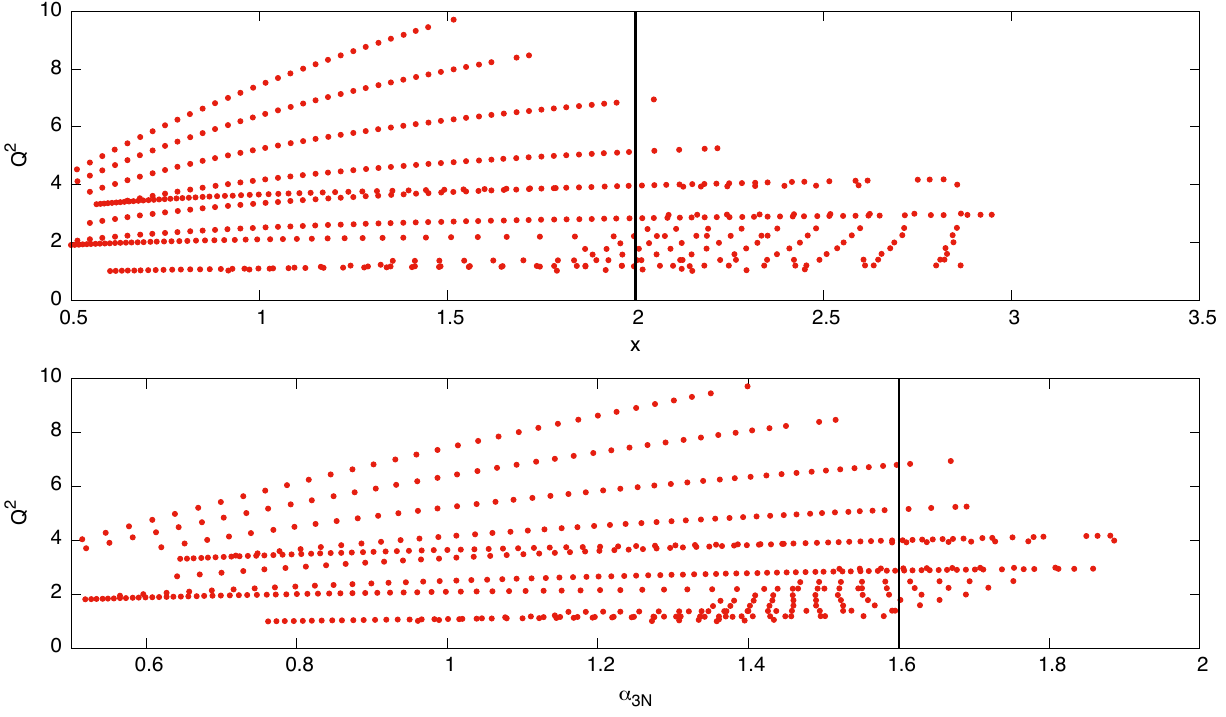}}
	\caption{Kinematic distribution of the world data set for $^3$He with $Q^2 > 1$: Q$^2$ versus $x$ (top) and Q$^2$ versus $\alpha_{3N}$ (bottom). Only data with $\alpha_{3N}> 1.6$ is used when considering 3N-SRC as indicated by the black vertical line in the bottom panel. The kinematic points above do not necessarily imply  corresponding data for other nuclei. It should be noted that at very large $x$ and $\alpha_{3N}$ the data have  large relative errors.}
      \label{AllData}
   \end{figure}

The data available to study   3N-SRCs are sorely limited.  $^3$He data (SLAC\cite{Rock:1981aa}, Hall B \cite{Kim1,Kim2},  Hall C \cite{Fomin2011} and Hall A \cite{Ye:2017mvo} at Jefferson Lab) provided  good agreement for the height of the 2N-SRC plateau 
 for $x<1.5<2.0$  yet there are significant disagreements in the $x>2$ region. These arise from the fact that the SLAC 
 and  data from Jefferson Lab's Hall A\cite{Ye:2017mvo} are at the lower limit of the range of $Q^2$ necessary to study 3N correlations 
 and the same is true for a fraction of the data from CLAS\cite{Kim1,Kim2}.   The reliability of the observed scaling in the  $x>2$ region for CLAS 
 data  was questioned in Ref.\cite{Higinbotham:2014xna}  which observed that the modest momentum resolution of the CLAS 
 detector in Hall B allows, when the cross sections are falling steeply with $x$, bin migration, in which events in a reconstructed $x$-bin 
 originated in a lower $x$-bin. To get a sense for paucity of the data, we show in Fig.~\ref{AllData} the kinematic extent of  all 
 published $^3$He data cited above as a scatter plot of $Q^2$ and $x$ (top)  and   $Q^2$ and $\alpha_{3N}$ (bottom). As it can be seen only a small fraction of the data satisfy the necessary condition of $Q^2\gtrsim 3$~GeV$^2$ and $\alpha_{3N}\gtrsim1.6$ as 
 indicated by the vertical line even though the large set of data to the right of the vertical line at $x=2$ in the top panel might suggest  otherwise.

 Experiment E02-109\cite{Arrington:2002ex,Fomin:2010ei,FominAIP,Fominpc,fominphd} was developed with the goal, in part, to provide precision ratios, in the 2N-SRC region, at large momentum transfer for a wide range of nuclei. The ratios, $({2\sigma_A\over A \sigma_D})$, at $Q^2\approx 2.75~\textrm{GeV}^2$ (at $x =1$) indicated scaling patterns expected for 2N-SRC region\cite{Fomin2011}. The heights of the plateaus at $x > 1.5$ scale approximately with A\cite{Fomin2011} (see Fig.~\ref{2NSRC-Fomin}) and  have been related to the parameter $a_2(A,Z)$ (Eq.~\ref{a2}) characterizing the probability of finding 2N-SRCs in nucleus $A$ relative to the deuteron.   

\begin{figure}[htbp]
      \centering
           \includegraphics[width=\textwidth]{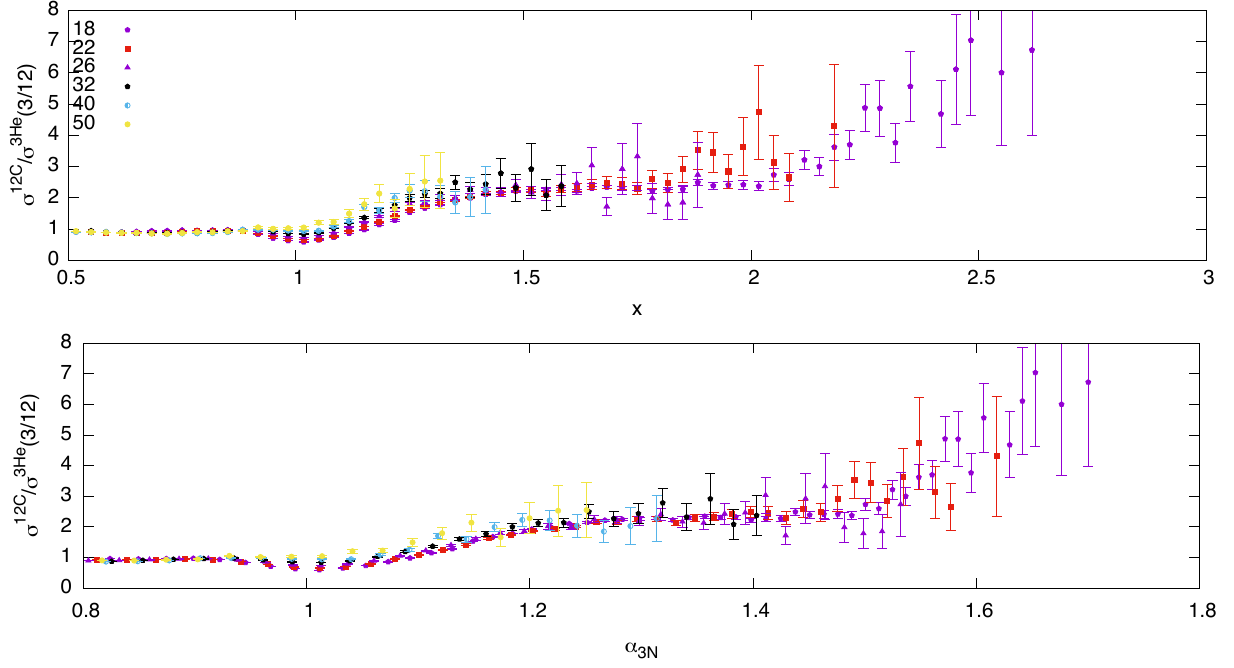} 
      \caption{The $x$ and $\alpha_{3N}$ dependences of the per-nucleon ratios of $^{12}$C/$^3$He for different angles with $Q^2$  ranging from $2.5-7.5$~GeV$^2$ (at  $x=1$) against $x$ (top) and $\alpha_{3N}$. Only data with relative errors less than 0.5 are shown.}
     \label{x_alpha3N}
   \end{figure}

To check whether $\alpha_{3N}$ description of the data results in the 
factorization implicit in Eq.~(\ref{crosssection}), in Fig.~\ref{x_alpha3N} we compare
the $x$ and $\alpha_{3N}$ dependences of ${3\sigma_A\over A \sigma_{^3He}}$ for all available $Q^2$ from Refs.~\cite{Fomin2011,fominphd} for $A=12$. 
 As the comparison shows the $Q^2$ spread of the data is significantly reduced once the ratios are evaluated in terms of $\alpha_{3N}$ which absorbs part of the $Q^2$ dependence. The plateau in the region  $1.3 < \alpha_{3N}< 1.5$ manifests the dominance of 2N-SRCs  corresponding, here, to internal nucleon momenta in the range of $300-600$~MeV/c.   The plateaus arising from 2N-SRCs in the ratios  as a function of  $\alpha_{3N}$ (similar to  $\alpha_{2N}$)   can be seen as following from the fact (see Fig.~\ref{al_x}) that numerically $\alpha_{2N}$ and $\alpha_{3N}$ have small differences at $Q^2 > 2$~GeV$^2$ and $x<1.8$.  The observation of 2N-SRCs in terms of  $\alpha_{3N}$ is important for verifying the conjecture  (Eq.~\ref{R3_R2sq_simp})) that a plateau, if observed, in the 3N-SRC region  should be proportional to  $({a_{2}(A)\over a_2(^3\textrm{He})})^2$. 

As Fig.\ref{x_alpha3N} shows only measurement at 18$^\circ$, (in which $Q^2\simeq 2.5$ (GeV/c)$^2$ at the quasielastic peak, growing to $Q^2 \simeq 3$ (GeV/c)$^2$ at $x = 2.9$) reaches to the region  $\alpha_{3N}\simeq 1.6$ where one expects the onset of 3N SRCs. 
It is intriguing that  as the  lower panel of the figure shows  at $\alpha_{3N}>1.6$ the ratios indicate  possible onset of  the scaling.
In the  further discussions, except where explicitly indicated our analysis of \cite{Fomin2011} is limited to these  data  set.

Problems arose when constructing  the $^3$He  cross section between $x = 2.68$ and $x = 2.85$  $(1.6 \leq \alpha_{3N} \leq 1.8)$ due to difficulties with the subtraction of the walls of the Aluminum target cell containing the $^3$He. The limited vertex resolution of the spectrometer  made it impossible to isolate electrons that scattered from the walls of the relatively small diameter  (4~cm) target cell. This and the fact that $\sigma^{\rm{Al}} \gg \sigma^{^3\rm{He}}$ at large $x$ as $\sigma^{^3\rm{He}}$ must go to 0 at its kinematic limit, $x=3$. resulted in a set of negative cross sections  in  three bins at large $x$  interlaced with other bins in which cross sections were consistent with zero with large relative errors.

\begin{figure}[htbp]
      \centering
           \includegraphics[width=\textwidth]{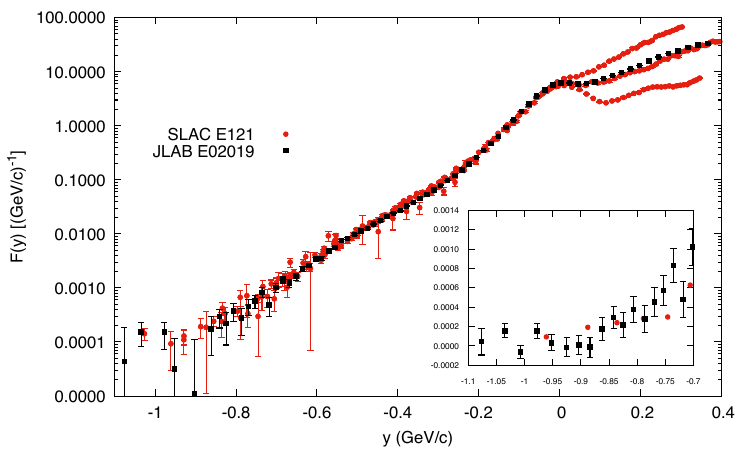} 
      \caption{$F(y)$ plotted against $y$ for $^3$He data from Ref.~\cite{Day:1979bx,Rock:1981aa,Fomin2011}. The inset shows  $F(y)$ for $ -0.7 > y > -1.1$. This corresponds to the region of interest for 3N-SRCs, $\alpha_{3N} \simeq 1.6$ (at $y = -0.7$) to $\alpha_{3N} \simeq 1.8 $ (at $y = -1.1$). The selected SLAC data shown here have $1< Q^2 < 4$ GeV/c$^2$. There is good agreement between the SLAC and Jefferson Lab data.} 
     \label{yscale}
   \end{figure}

In contrast, the data in the region below $x<2.5$ are of excellent quality  with  small errors.  As expected a $y$-scaling analysis\cite{Sick:1980ey,Day:1990mf} of the E02-019 data  found it to be in good agreement with the  SLAC data\cite{Day:1979bx,Rock:1981aa} from $y=0$ (top of the quasielastic peak) to $y \simeq -1$ (GeV/c). In Fig.~\ref{yscale} we plot the scaling function $F(y)$ against $y$ with the inset showing (in a linear scale) the region $ -1.1 < y < -0.7$ and where the negative values of $F(y)$ arise from the negative $^3$He cross sections mentioned above. 
Despite the negative $^3$He cross sections the ratio, $\frac{^4He}{^3He}$, over the entire x-region from E02-019 were formed and published in Ref.~\cite{Fomin2011} by making use of the following procedure. First,  an inverted ratio, $\frac{^3He}{4He}$, was formed and then, for the region of $x \geq 1.15$, the data was rebinned by combining three bins into one taking care of the error propagation. Subsequently the data in the inverted ratio that had error bars falling below zero were moved along a truncated gaussian, such that the lower edge of the error bar was at zero.       The result was then inverted to give the ratio $^4$He/$^3$He shown in Figure 3 of Ref.~\cite{Fomin2011} and as the  triangles in  Fig.~\ref{ratio4to3} below. The use of a truncated gaussian gave rise to the asymmetric error bars seen  in the ratio. A limitation of this approach is that it would have to be repeated for every nucleus when forming  $\frac{\sigma^A}{\sigma^{3He}}$.

As an alternative to the procedure of Ref.\cite{Fomin2011}  we have used the following approach\cite{Sargsian:2019joj} to avoid the  problematic $^3$He data of Refs.\cite{Arrington:2002ex,Fomin2011,Fomin:2010ei} in the 3N-SRC region. We fit the y-scaling function $F(y)$, derived from the SLAC $^3$He data between $x = 2.68$ and $x = 2.85$  $(1.6 \leq \alpha_{3N} \leq 1.8)$. The fit was of the form $F(y) = a \exp({-by})$. We were then able to replace, point by point,  the $^3$He cross sections from E02019 where its central value was negative or its error bar  fell below zero,  through the following:  $\frac{d^{2}\sigma}{d\Omega dE'} = \sum\limits_{eN}\sigma_{eN} \cdot K \cdot F(y)$, where K is a kinematic factor and  $\sigma_{eN}$ is the elementary electron-nucleon cross section. The  absolute  error of the E02019 data set \cite{Arrington:2002ex,Fomin2011,Fomin:2010ei} was kept rather than the smaller errors from the fit.  The fit parameters  are $a =0.296$ and $b = 8.241$. Note that, a  similar  approach was   used in Ref.~\cite{FSDS}, where the first evidence of 2N-SRCs through cross section ratios in inclusive scattering were revealed.  Subsequently those results were confirmed by  precision studies\cite{Kim1,Kim2,Fomin2011} in which the heavy and light cross section data  were measured in single experiment.

\begin{figure}[th]
\includegraphics[width=0.8\textwidth]{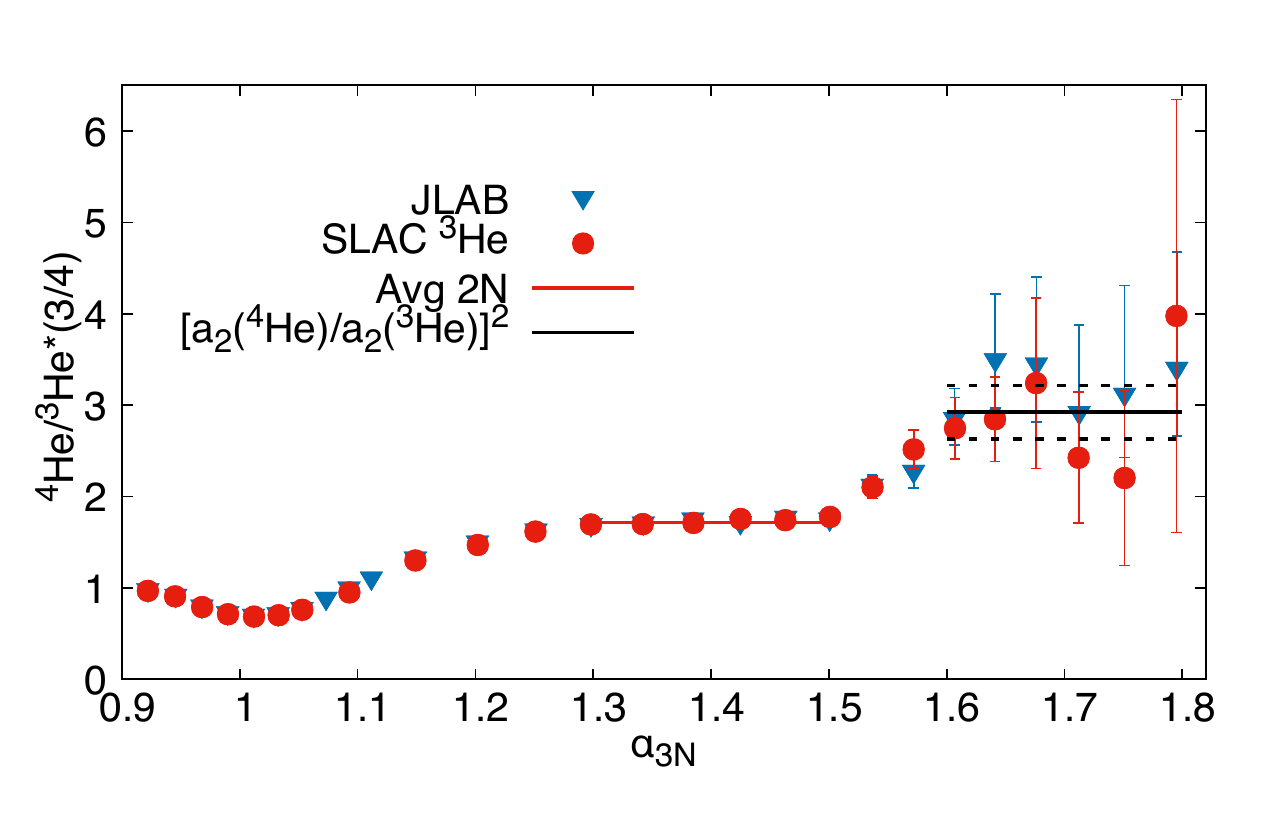}
\vspace{-0.5cm}
\caption{The $\alpha_{3N}$ dependence of the inclusive cross section ratios for $^4$He to  $^3$He, triangles - JLAB data~\cite{Fomin2011,Fomin:2010ei}, 
circles - ratios  when using a parameterization of SLAC $^3$He cross sections~\cite{Day:1979bx,Rock:1981aa}. 
The horizontal line at $1.3\le \alpha_{3N}<1.5$ identifies the magnitude of the 2N-SRC plateau. 
The line for $\alpha_{3N}>1.6$ is Eq.(\ref{R3_R2sq_simp}) with a $10\%$ error introduced to account for the systematic uncertainty in $a_2(A,Z)$ parameters across
all measurements. The data correspond to $Q^2\approx 2.5$~GeV$^2$ at $x=1, \alpha_{3N} =1$. The figure is adapted from 
Ref.\cite{Sargsian:2019joj}.}
\label{ratio4to3}
\end{figure}

Fig.~\ref{ratio4to3} presents the results  for the cross section ratios obtained from the   approaches described above;  the one adopted in Ref.\cite{Fomin2011} (blue triangles) and other (red circles) in which the scaling function $F(y)$ is used to reconstruct  cross sections between  $x = 2.68$ and $x = 2.85$ $(1.6 \leq \alpha_{3N} \leq 1.8)$. While both give similar results we consider  the replacement of the  problematic data points as a best alternative procedure of  Ref~\cite{Fomin2011}  in part because it allows a consistent treatment of the ratios for all $A $. 

\subsection{Studies of systematics}

\begin{figure}[htbp]
\begin{center}
\vspace{-0.4cm}
\includegraphics[width=0.45\textwidth]{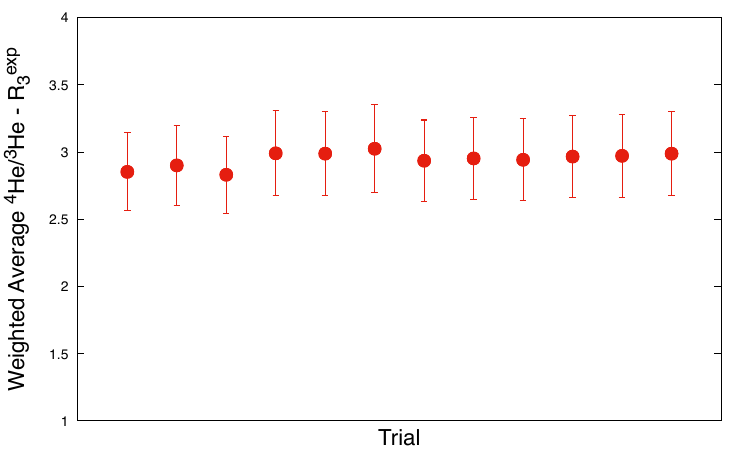} 
\includegraphics[width=0.45\textwidth]{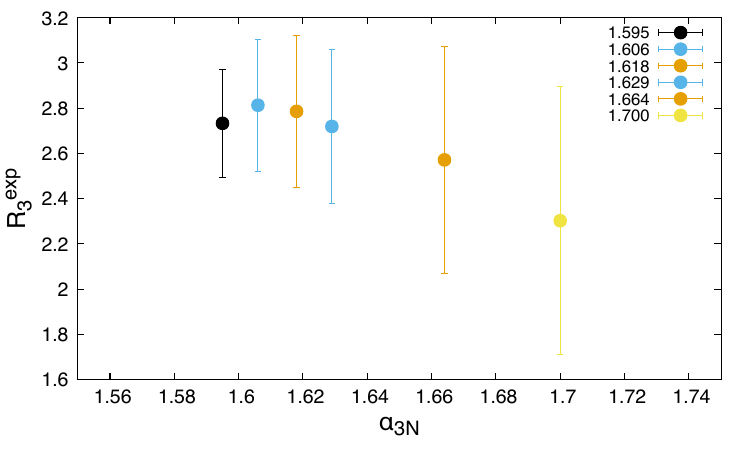}
\caption{Left: Sensitivity of Ratio, $\frac{^4He}{^3He}$ to  the 
procedure of replacing  the anomalous $^3$He data for 12 different trials; 
four of the data set used to form $F(y)$, including  
three sets for which fraction of the data have been replaced. 
Right:   Sensitivity of the weighted average of $\frac{^4He}{^3He}$ in the 3N-SRC region 
on the lower limit of $\alpha_{3N}$.
The results shown in Fig.~\ref{R2andR3anda2}(a) remain unchanged within errors which 
grow with a larger $\alpha_{3N}>1.6$ limit.  In all cases (here and our final results) we restricted the upper limit such that $W_{3N}$ was at least 50 MeV less than the elastic limit 
$\alpha_{3N} \le 1.75$, $x \le 2.85$ and $ y \ge -0.92$
so as to avoid, as much as possible, FSI.}
\label{SysStudies}
\end{center}
\end{figure}

We have worked to evaluate the sensitivity of the procedure above to obtain the $\frac{^4He}{^3He}$  ratios, as measured by $R_3^{exp}$ 
(Eq.(\ref{R3nsrc})), in multiple ways. We varied the  both the data sets used in the fit to F(y) and the fraction of the data in the fit range, $ -1.08 \ge y \le -0.84$, that is replaced.  We made fits to $F(y)$ built from 4 different data sets (1) the  SLAC data only, (2) the  JLAB data only, (3) both the JLAB and the SLAC data, and finally (4) the JLAB data absent its negative values.  In addition, using the 4 fits to $F(y)$, we examined 3 variations of the fraction of the JLAB data set replaced by cross sections from the $F(y)$ fits: all the data in the fit range; just the 6 bins in the fit range where the data or its error bar went below zero; and only the data where the  cross section values were negative. We found that these 12 variations for the $^3$He cross sections  resulted in weighted averages in the 3N region that agreed easily within their error bars. See Figure~\ref{SysStudies}. We  show  on the right hand side of Figure~\ref{SysStudies} the dependence of $R_3^{exp}$ on the lower limit in $\alpha_{3N}$ when taking the weighted average. As can be seen the result is not strongly dependent on $\alpha_{3N}$ and the error bar increases due to the worsening statistics.

\medskip

Going back to  Fig.~\ref{ratio4to3}  we notice that the plateau  due to 2N-SRCs is clearly visible   for $1.3 \le \alpha_{3N} \le 1.5$.   In this region $\alpha_{3N}\approx \alpha_{2N}$, where $\alpha_{2N}$ is the LC momentum fraction of the nucleon in  the  2N-SRC. Because of this, we refer to   the magnitude of this plateau as $R_2(A,Z)$ defined in Eq.(\ref{R2_def}).

   \begin{figure}[!htbp]
     \centering
 \includegraphics[width=0.9\textwidth]{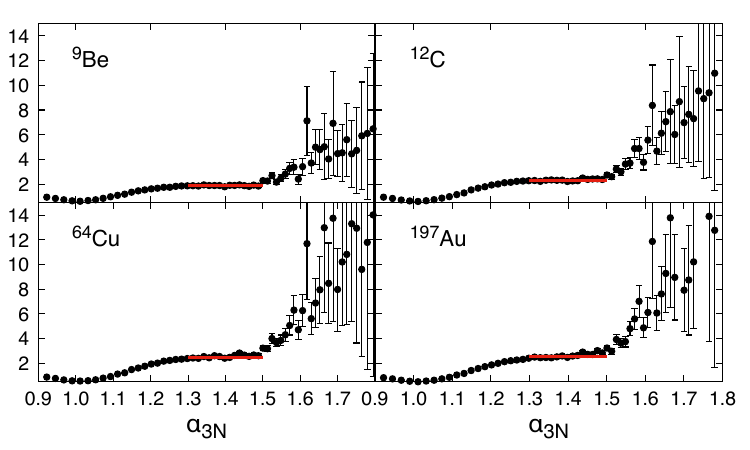}
   \caption{Per-nucleon cross section ratios for  $^9$Be, $^{12}$C, $^{64}$Cu, $^{197}$Au  to  $^3$He. Horizontal lines 
    indicating $a_2(A)\over a_2(^3He)$ in the 2N-SRC  region.}
\label{4panel3Heratios}
   \end{figure}
  
   The horizontal line in the region of $1.3 \le \alpha_{3N} \le 1.5$ is  given by the right hand side of  Eq.~(\ref{R2_def}), in which  the values of 
 $a_2(^3He)$ and  $a_2(A)$  are taken from the last column of Table II in Ref.~\cite{Arrington:2012ax}, an average of the SLAC, JLAB Hall C and JLAB Hall B results. The magnitude of the  horizontal solid line in the region of  $1.6 \le \alpha_{3N} \le 1.8$, is  the prediction 
 of $R_{3N}(A,Z)\approx R_{2N}^2(A,Z)$  which was explained in the previous section (Eq.(\ref{R3_R2sq_simp})).  We assigned a $10\%$ error to this prediction (dashed lines) related to the uncertainty of 
$a_2(A,Z)$ magnitudes across different measurements.

  With the same $^3$He cross sections in  Fig.~\ref{4panel3Heratios} we evaluated 
 ratios of cross sections, ${3\sigma^A \over A\sigma^{3He}}$ for the nuclei ($^4$He, $^9$Be, $^{12}$C, $^{64}$Cu and $^{197}$Au).  
 Additionally in this figure,   we evaluated the magnitudes of ${a_2(A)\over a_2(^3He)}$     (taken from Ref.~\cite{Fomin2011}) which are indicated by  horizontal lines for  $ 1.3 \leq \alpha_{3N} \leq 1.5$ where the plateau due to 2N-SRCs is observed. 
  
\begin{figure}[!htbp]
\includegraphics[width=0.45\textwidth]{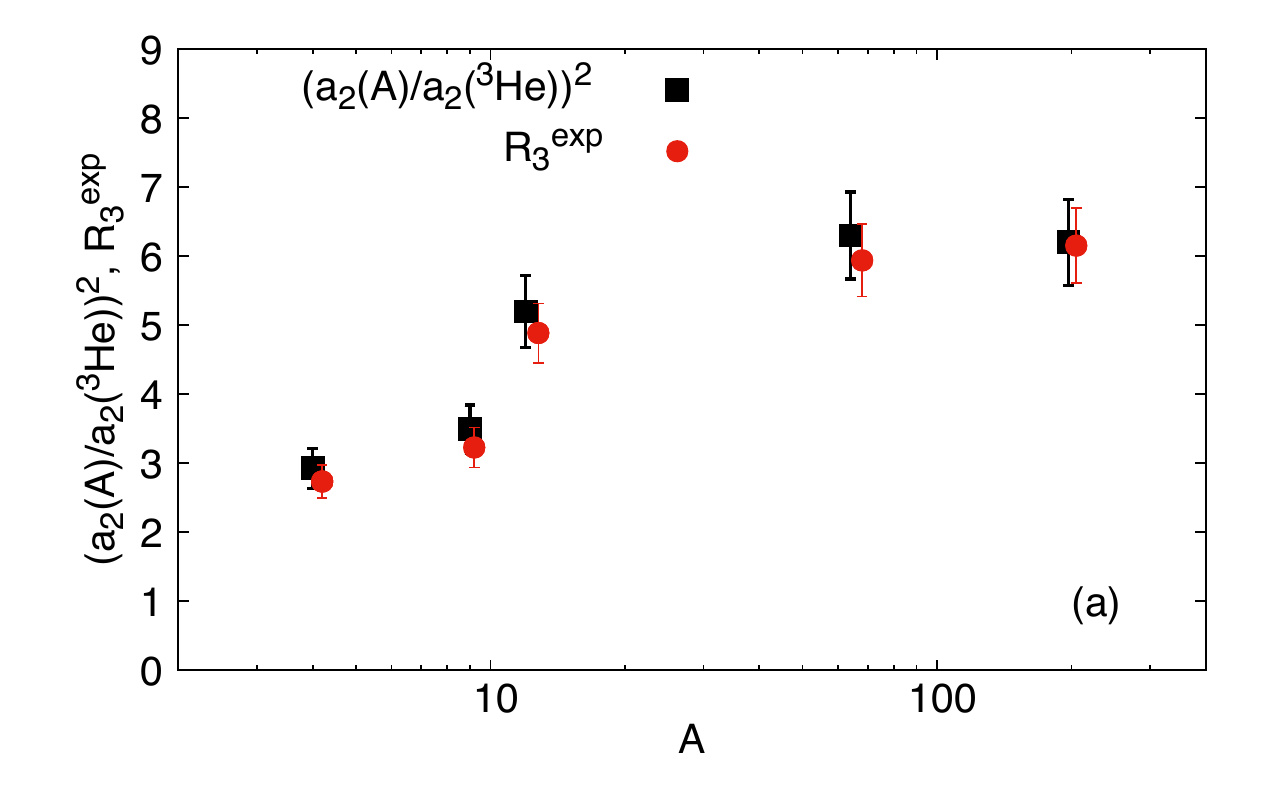}
\includegraphics[width=0.45\textwidth]{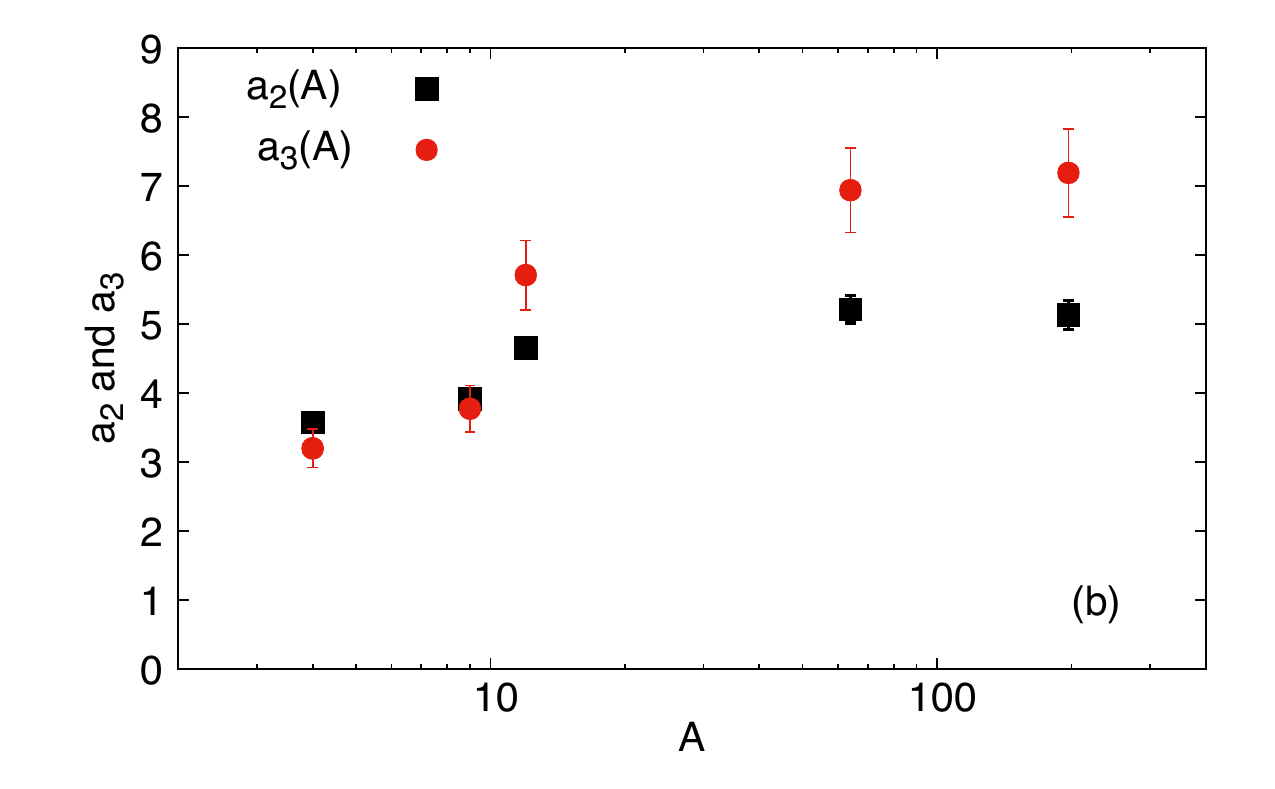}
\caption{(a) The $A$ dependence of the experimental evaluation of $R_3$ compared with the 
prediction of Eq.\ref{R3_R2sq_simp}. (b) The $A$ dependence of 
$a_3(A,Z)$ parameter compared to $a_2(A,Z)$ of Ref.\cite{Fomin2011}. }
\label{R2andR3anda2}
\end{figure}

  As can be seen  from these figures, despite large errors,  the data (similar to  Fig.\ref{ratio4to3}) indicate a strong enhancement in the ratio, $R_3(A)$ as soon as $\alpha_{3N}\gtrsim 1.6$ which 
 are in qualitative  agreement with the  prediction of Eq.(\ref{R3_R2sq_simp}).   
  To test quantitatively the prediction of  Eq.~(\ref{R3_R2sq_simp}), in   Fig.~\ref{R2andR3anda2}(a) 
  we evaluated the   weighted  average of $R^{exp}_3(A,Z)$ for $\alpha_{3N}> 1.6$ and compared them with the magnitude of  
$({a_2(A,Z)\over a_2({3He})})^2$ in which $a_2(A,Z)$'s are taken from  Ref.~\cite{Arrington:2012ax}.
In these evaluations $^3$He cross sections were taken from the $F(y)$ fit to the SLAC data.
 Numerical data of Fig.~\ref{R2andR3anda2}  are presented also  Table~\ref{a2R3a3}. 
The comparison in Fig.~\ref{R2andR3anda2}(a) shows   good agreement with the prediction of Eq.(\ref{R3_R2sq_simp}) for the full  range of nuclei. 
We investigated the sensitivity of the weighted average of $R_3(A,Z)$ on the lower limit of $\alpha_{3N}$ (before rebinning) 
and found that the results shown in Fig.~\ref{R2andR3anda2}(a) remain unchanged within errors which 
grow with a larger $\alpha_{3N}>1.6$ cut.

  The agreement presented in Fig.\ref{R2andR3anda2}(a) represents the strongest evidence yet for  the presence of 3N-SRCs.
If it  is indeed due to the onset of  3N-SRCs then one can use the measured $R^{exp}_3$ ratios and Eq.(\ref{R3_a3}) to extract the $a_3(A,Z)$ 
parameters characterizing the 3N - SRC probabilities in the nuclear ground state. The estimates of $a_3(A,Z)$ and comparisons with 
$a_2(A,Z)$ are  given in Fig.\ref{R2andR3anda2}(b) (see also Table~\ref{a2R3a3}). 
These comparisons show a faster rise for $a_3(A,Z)$  with $A$, consistent with the expectation 
of the  increased sensitivity of 3N-SRCs to  the local  nuclear density\cite{srcrev}.  If this result is verified in the future with better quality 
data and a wider range of nuclei  then the evaluation of the parameter $a_3(A,Z)$ as a function of nuclear density and 
proton/neutron asymmetry together with $a_2(A,Z)$  can provide an important theoretical input for the  exploration of the dynamics of super dense nuclear matter (see e.g. \cite{Ding:2016oxp}).

\begin{table}[ht]
\centering
\caption{Numerical values a$_2$\cite{Arrington:2012ax}, R$_2$ (Eq.~\ref{R2_def}), R$_3^\textrm{exp}$ (the weighted average in the 3N region)  and  $a_3$  calculated from Eq.~\ref{R3_a3}.}
\vspace{0.2cm}
\begin{tabular}{|c|c|c|c|c|}\hline
A   & a$_2$    & $R_2$       &  $R_3^{\textrm{exp}}$ &  a$_3$ \\ \hline
3   & 2.13 $\pm 0.04$ &  1  &   NA                     &    NA	   \\ \hline
4   & 3.57 $\pm 0.09$ & $1.68 \pm  0.03$  &  2.74 $\pm 0.24$  &  $3.20 \pm 0.28$  \\ \hline
9   & 3.91 $\pm 0.12$ & $1.84 \pm  0.04$  &  3.23 $\pm 0.29$  &  $3.77 \pm 0.34$  \\ \hline
12  & 4.65 $\pm 0.14$ & $2.18 \pm  0.04$   & 4.89 $\pm 0.43$   &  $5.71 \pm 0,50$   \\ \hline
64  & 5.21 $\pm 0.20$ & $2.45 \pm  0.04$   & 5.94 $\pm 0.52$   &  $6.94 \pm 0.77$   \\ \hline
197 & 5.13 $\pm 0.21$ & $2.41 \pm  0.05$  & 6.15 $\pm 0.55$   &  $7.18 \pm 0.64$   \\ \hline
\end{tabular}
\vspace{0.2cm}
\label{a2R3a3}
\end{table}

\section{ Summary and outlook} 
\label{sec7}

We determined the kinematic conditions for isolating 3N SRCs in inclusive $A(e,e^\prime)X$ reaction at large $x$. Based on the analysis of short range structure of $^3$He nuclei we expect that the dominant mechanism of 3N SRCs in inclusive processes is due to three-nucleon correlations, in which one  fast nucleon is balanced by two spectator nucleons with rather small  invariant  mass,
$2m_N\le m_S \lesssim 1.9$~GeV. Momenta of all three nucleons, however, exceed the characteristic Fermi momentum of the nucleus $k_{\textrm{F}}\sim 250$~MeV/c. We referred such correlations as type 3N-I SRCs.

We explain that due to the specific nature of the high momentum components of the nuclear wave functions, the momentum of the fast 
nucleon is not the optimal variable for the analysis since it does not allow the separation of 2N  and 3N~SRCs.  
In this respect the light-cone momentum fraction of  3N-SRC carried by the interacting nucleon, $\alpha_{3N}$  is more suitable 
and existing phenomenology indicates that the onset of the 3N-SRC dominance is expected at $\alpha_{3N} >1.6$. 
We derived the expression for  $\alpha_{3N}$ for inclusive $A(e,e^\prime)X$  processes and demonstrated that the 
$\alpha_{3N} \gtrsim1.6$ condition puts  a strong constraint on $Q^2$ of the reaction -  requiring $Q^2\gtrsim 3$~GeV$^2$.
Under these conditions we expect that the dominance of 3N-SRCs  will  lead to 
a  plateau for  per-nucleon inclusive cross section ratios  of heavy to light nuclei.  This will be in   addition
to the plateau  observed in the 2N-SRC region.

Furthermore, based on the $pn$ dominance in  2N-SRCs we predict that 3N-SRCs are generated through two
successive $pn$ short range interactions. Within   such  scenario we derived a quadratic 
relation between per nucleon ratios of nuclear and $^3He$ inclusive cross sections measured in the 2N-  ($R_2$) and 3N- ($R_3$) SRC 
regions: $R_3\approx R_2^2$. 

We analyzed the existing inclusive data under the 
above conditions and found an indication for  the onset of the plateau at 
$\alpha_{3N}>1.6$.  It is very intriguing that the magnitude of the plateau, $R_3$  is in agreement with predicted 
$R_3\approx R_2^2 \approx  ({a_2(A)\over a_2(^3He)})^2$  dependence.  This agreement allowed us 
to extract  per nucleon probabilities, $a_3(A,Z)$  of  finding 3N-SRCs in nuclei-A relative to the $^3$He nucleus.
  
\begin{figure}
     \centering
          \includegraphics[width=.75\textwidth]{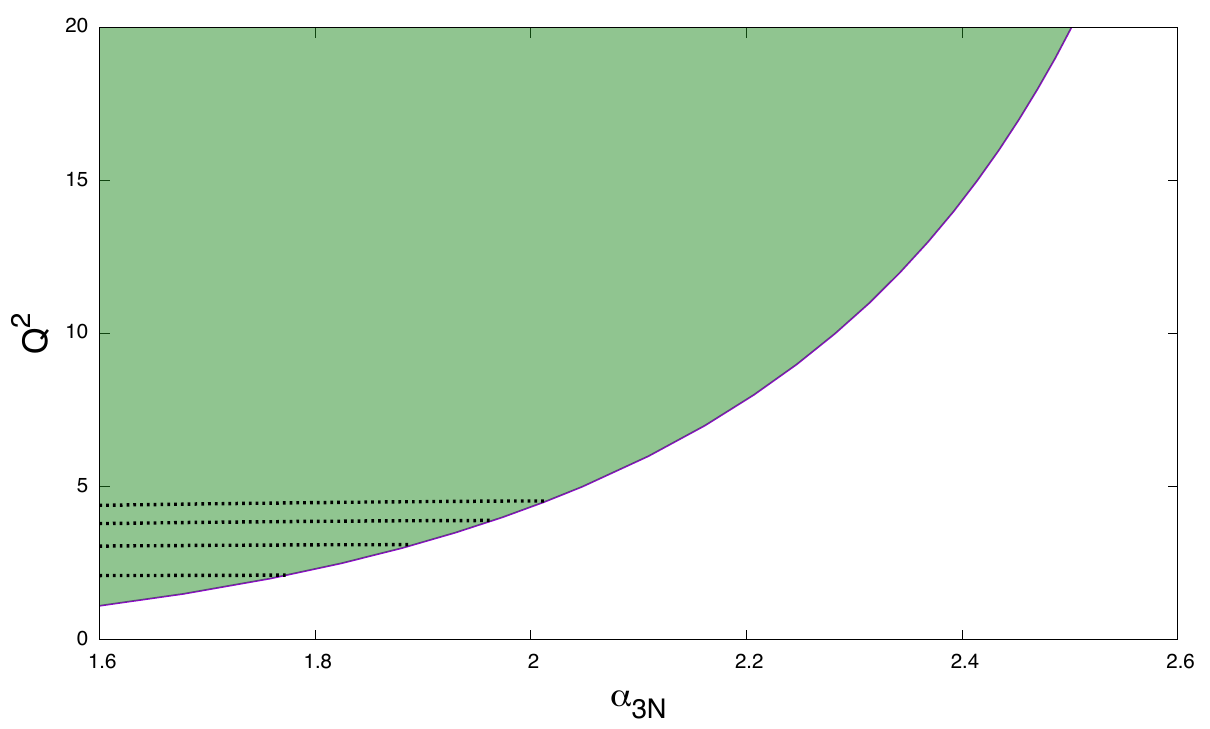}
   \caption{The $Q^2$ range necessary in order to isolate 3N-SRCs. Also shown is the kinematic extent of an upcoming 12 GeV experiment \cite{Arrington:2006xx}. }
\label{Q2_range}
   \end{figure}

The forthcoming experiments at Jefferson Lab will be able to significantly improve current experimental situation.  One important condition is that such experiments will be able to cover a larger $Q^2$ region.  As Fig.~\ref{Q2_range} shows an increase of $Q^2$ will significantly 
widen the range of the $\alpha_{3N}$ accessible by the experiment.  It is worth mentioning that at $Q^2\gtrsim 5$~(GeV/c)$^2$ one will be able to cross to the $\alpha_{3N} \ge 2$  region where one expects maximal contribution due to 3N SRCs.

It is with anticipation that we await the  running and analysis  of Jefferson Lab's E1206-105\cite{Arrington:2006xx} experiment
which has multiple goals:  to measure cross sections  1) from light nuclei to compare to ab-initio calcuations and to study FSI, 2) from nuclei at low and moderate $Q^2$ with a range of $p-n$  asymmetries  in order to look for isospin dependence in the per-nucleon ratios, 
3) at moderate $Q^2$ and large $x$ to search for definitive evidence to 3N~SRCs and finally 4) at very large $Q^2$ to look for the transition from quasielastic to deep inelastic scattering  from nuclei as part of an effort to extract nuclear parton distribution functions at $x>1$.  The drawn lines in Fig.~\ref{Q2_range} indicate the tentative range in $Q^2$ and $\alpha_{3N}$ which will be part of the goal of 
this experiment in  studying 3N~SRCs.
 
Finally,  the discussed in this work type 3N-I SRCs correspond to those states in superdense nuclear matter in which no inelastic transition 
took place in the intermediate states. To investigate type 3N-II SRCs that are sensitive to irreducible 3N nuclear forces containing inelastic 
transitions  one will need studies of semi-inclusive processes in which nucleons from 3N-SRCs are detected in coincidence with scattered electron.
 In particular, it  would be instructive to compare production from  3N system in different isospin states: like three  protons, where contribution of the repulsive core is enhanced, and 2p+n state in which the attraction dominates.

 {\bf Acknowledgments:} 
 This work was supported in part by the DOE Office of Science, Office of Nuclear Physics, contracts DE-FG02-96ER40950 (DBD), 
 DE-FG02-93ER40771 (MIS), and  DE-FG02-01ER41172 (MSS).

%%%%%%%%%%%%%%%%%%%%%%%%%%%%%%%%%%%%%%%%%%%%%%%%%%%%%%%%%%%%%%%%%%%%%%%%%

\end{document}